\pdfoutput=1
\documentclass[12pt, letterpaper]{article}
\usepackage[left = 1in, right = 1in, top = 1in, bottom = 1.2in]{geometry}
\usepackage{amsfonts, amssymb, amsmath}
\usepackage{array}
\usepackage{booktabs}
\usepackage{natbib}
\usepackage{graphicx}
\usepackage{multirow}
\usepackage{enumitem}
\usepackage{tikz}
\usetikzlibrary{patterns}

\let\E\undefined
\def\E{\mathbb{E}}
\let\var\undefined
\DeclareMathOperator{\var}{Var}
\let\bias\undefined
\def\bias{\mathrm{bias}}
\def\pr{\mathbb{P}}
\let\T\undefined
\def\T{\mathrm{\scriptscriptstyle T}}
\DeclareMathOperator*{\argmax}{arg\,max}
\def\zero#1{0_{\,#1}}
\let\logit\undefined
\DeclareMathOperator{\logit}{logit}
\let\expit\undefined
\DeclareMathOperator{\expit}{expit}
\newcommand{\RomanNum}[1]{\MakeUppercase{\romannumeral #1\relax}}

\newcommand{\cc}[1]{\multicolumn{1}{c}{#1}}
\def\wt{\widetilde}
\def\wh{\widehat}
\def\wb{\overline}
\def\lb{\left\lbrace}
\def\rb{\right\rbrace}

\renewcommand{\baselinestretch}{1.6}

\def\loglikgamma{\ell_{\mathrm{t}}}
\def\loglikbeta{\ell_{\mathrm{o}}}

\def\extgamma{\Phi}
\def\extbeta{\Psi}

\def\ifgamma{\varphi_\gamma}
\def\ifbeta{\varphi_\beta}
\def\ifR{\varphi_R}
\def\ifRhat{\wh{\varphi}_R}

\def\ifRa{\varphi_{R1}}
\def\ifRahat{\wh{\varphi}_{R1}}
\def\ifRb{\varphi_{R2}}
\def\ifRbhat{\wh{\varphi}_{R2}}

\usepackage{algorithm2e}
\usepackage{placeins}

\title{Using Decision Lists to Construct Interpretable and Parsimonious Treatment Regimes}

\author{
  Yichi Zhang,
  Eric B. Laber,
  Anastasios Tsiatis,  
  and Marie Davidian\\ 
North Carolina State University}

\date{}

\begin{document}

\maketitle

\begin{abstract}
  A treatment regime formalizes personalized medicine as a function
  from individual patient characteristics to a recommended treatment.
  A high-quality treatment regime can improve patient outcomes while
  reducing cost, resource consumption, and treatment burden.  Thus,
  there is tremendous interest in estimating treatment regimes from
  observational and randomized studies.  However, the development of
  treatment regimes for application in clinical practice requires the
  long-term, joint effort of statisticians and clinical scientists.
  In this collaborative process, the statistician must integrate
  clinical science into the statistical models underlying a treatment
  regime and the clinician must scrutinize the estimated treatment
  regime for scientific validity.  To facilitate meaningful
  information exchange, it is important that estimated treatment
  regimes be interpretable in a subject-matter context.  We propose a
  simple, yet flexible class of treatment regimes whose members are
  representable as a short list of if-then statements.  Regimes in
  this class are immediately interpretable and are therefore an
  appealing choice for broad application in practice.  We derive a
  robust estimator of the optimal regime within this class and
  demonstrate its finite sample performance using simulation
  experiments.  The proposed method is illustrated with data from two
  clinical trials.

\end{abstract}

\section{Introduction}
Treatment regimes formalize clinical decision making as a function
from patient information to a recommended treatment.  Proponents of
personalized medicine envisage the widespread clinical use
of evidence-based, i.e., data-driven, treatment regimes.  The potential
benefits of applying treatment regimes are now widely recognized.  By
individualizing treatment, a treatment regime may improve patient
outcomes while reducing cost and the consumption of resources.  This is
important in an era of growing medical costs and an aging global
population.  However, the widespread integration of data-driven
treatment regimes into clinical practice is, and should be, an
incremental process wherein: (i) data are used to generate hypotheses
about optimal treatment regimes; (ii) the generated hypotheses are
scrutinized by clinical collaborators for scientific validity;
(iii) new data are collected for validation and new hypothesis
generation, and so on.  Within this process, it is crucial that estimated
treatment regimes be interpretable to clinicians.  
Nevertheless, optimality, not
interpretability, has been the focal point in the statistical
literature on treatment regimes.
   
A treatment regime said to be optimal if it maximizes the expectation
of a pre-specified clinical outcome when used to assign treatment to a
population of interest.  There is a large literature on estimating
optimal treatment regimes using data from observational or randomized
studies.  Broadly, these estimators can be categorized as
regression-based or classification-based estimators.  
Regression-based
estimators model features of the conditional distribution of the
outcome given treatment and patient covariates.  Examples include
estimators of the regression of an outcome on covariates, treatment,
and their interactions  
\citep[e.g.,][]{su2009subgroup, qian2011performance, tian2014simple},
and estimators of point treatment effects given covariates
\citep[e.g.,][]{robins1994correcting, 
vansteelandt2014structural}.   Regression-based methods rely on correct
specification of some or all of the modeled portions of the
conditional distribution of the outcome.  Thus, a goal of many
regression-based estimators is to ensure correct model specification
under a large class of generative models
\citep[][]{zhao2009reinforcement, qian2011performance,
  moodie2013q-learning, laber2014interactive, taylor2014reader}.
However, as flexibility is introduced into the model, interpretability
tends to diminish, and in the extreme case the estimated treatment
regime becomes an unintelligible black box.

% Examples include
% $Q$-learning \citep[][]{watkins1992q-learning,
% murphy2005generalization, zhao2009reinforcement,
% chakraborty2010inference, qian2011performance, goldberg2012q-learning,
% moodie2013q-learning, laber2014interactive, laber2014set} and $g$-estimation in structural nested mean models
% \citep[][]{murphy2003optimal, robins2004optimal,
%   moodie2007demystifying, henderson2009regret, schulte2014q-and-a,
%   tian2014simple}.  Regression-based methods rely on correct
% specification of some or all of the modeled portions of the
% conditional distribution of the outcome.  Thus, a goal of many
% regression-based estimators is to ensure correct model specification
% under a large class of generative models
% \citep[][]{zhao2009reinforcement, qian2011performance,
%   moodie2013q-learning, laber2014interactive, taylor2014reader}.
% However, as flexibility is introduced into the model, interpretability
% tends to diminish, and in the extreme case the estimated treatment
% regime becomes an unintelligible black box.

Classification-based estimators, also known as policy-search or
value-search estimators, estimate the marginal mean of the outcome for
every treatment regime within a pre-specified class and then take the
maximizer as the estimated optimal regime.  Examples include marginal
structural mean models \citep[][]{robins2008estimation,
  orellana2010dynamic}; robust marginal mean models
\citep[][]{zhang2012robust}; and outcome weighted
learning \citep[][]{frank, zhao2012estimating,
  zhao2014doubly}.  Classification-based estimators often rely on fewer
assumptions about the conditional distribution of the outcome given
treatment and patient information and thus may be more robust to model
misspecification than regression-based estimators
\citep[][]{zhang2012robust, frank}.  Furthermore,
because classification-based methods estimate an optimal regime
within a pre-specified class, it is straightforward to impose
structure on the estimated regime, e.g., interpretability, by
restricting this class.  We use robust marginal mean models with a
highly interpretable yet flexible class of regimes to estimate a
high-quality regime that can be immediately understood by clinical and
intervention scientists.

\begin{figure}
\renewcommand{\baselinestretch}{1.1}\fontsize{11}{12}\selectfont
\centering
\includegraphics[scale = 0.7]{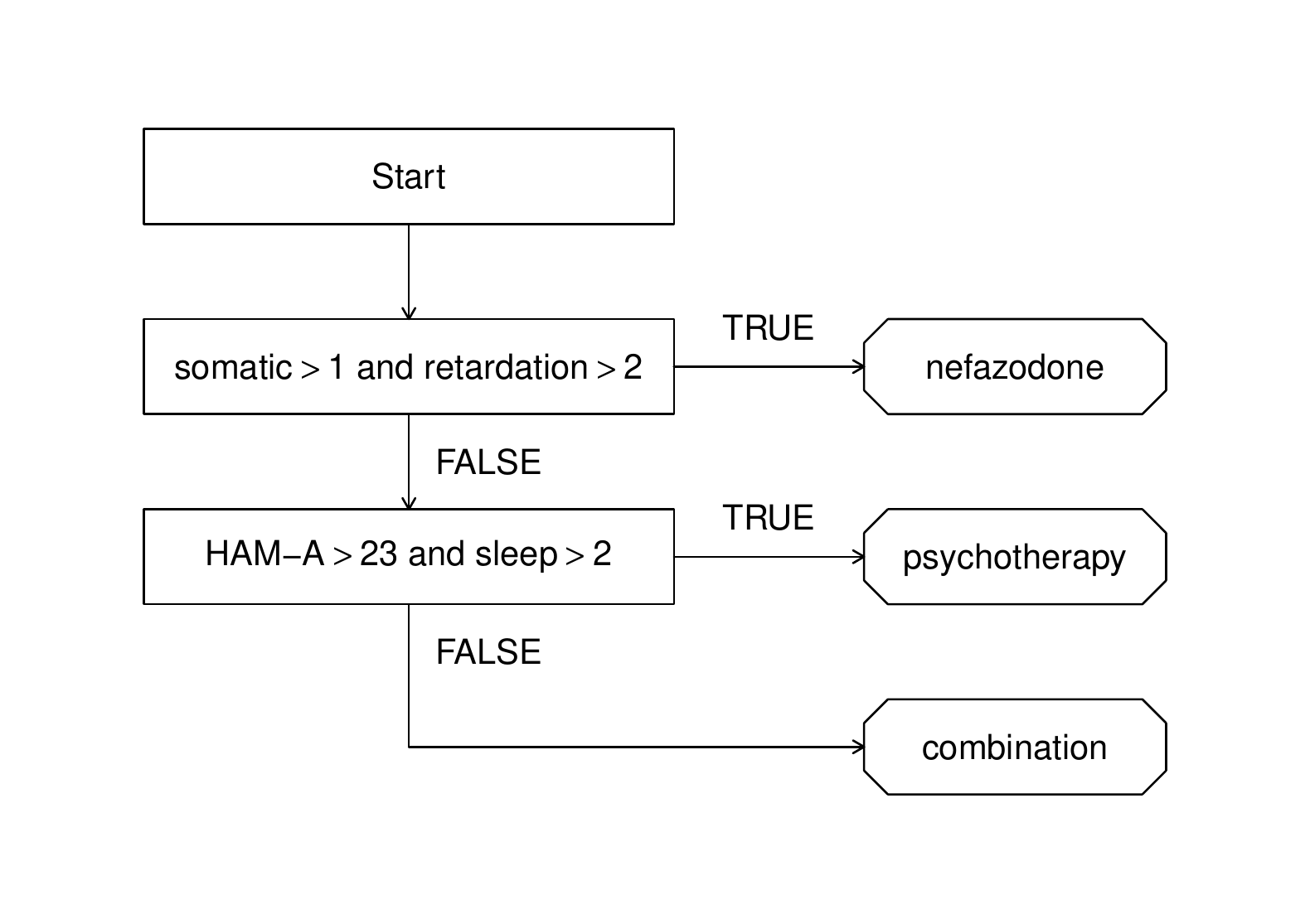}
\vspace*{-4ex}%
\caption{Estimated decision list for treating patients with chronic depression.}
\label{fig:example}
\end{figure}

To obtain an interpretable and parsimonious treatment regime, we
use the concept of decision list, which was developed in the
computer science literature for representing flexible but
interpretable classifiers \citep{rivest1987learning, clark1989cn2,
  marchand2005learning, letham2012building, wang2014falling};
see \cite{freitas2014comprehensible} for a recent position 
paper on the importance of interpretability in predictive
modeling and additional
references on interpretable classifiers.  
As a treatment regime, a
decision list comprises a sequence of ``if-then'' clauses that map
patient covariates to a recommended treatment.
Figure~\ref{fig:example} shows a decision list for patients with
chronic depression (see Section~4.2).  This decision list recommends
treatments as follows: if a patient presents with somatic anxiety
score above $1$ and retardation score above $2$, the list
recommends nefazodone; otherwise, if the patient has Hamilton anxiety
score above $23$ and sleep disturbance score above $2$, the list
recommends psychotherapy; and otherwise the list recommends nefazodone
+ psychotherapy (combination).  Thus, a treatment regime represented
as a decision list can be conveyed as either a diagram or text and is
easily understood, in either form, by domain experts.  
Indeed, decision lists have frequently been used to display 
estimated treatment regimes 
 \citep[][]{shortreed2011informing, 
moodie2012q, shortreed2014multiple, treeBased} or to
describe theory-based, i.e., not data-driven, treatment regimes 
\citep[][]{shiffman1997representation, 
marlowe2012adaptive}.
  
% Although \citet{shortreed2011informing, shortreed2014multiple} showed how to build regimes in an if-then format 
% upon $Q$-learning by enumerating levels of categorical covariates and 
% eliminating continuous covariates irrelevant to treatment decisions,
% this ad-hoc approach fails if most of the useful covariates are continuous or the $Q$-function is misspecified.
% Thus, we propose a more direct and robust approach in this paper.
Another important attribute of a decision list is that it ``short circuits''
measurement of patient covariates; e.g., in Figure~\ref{fig:example},
the Hamilton anxiety score and sleep disturbance score do not need to
be collected for patients with somatic anxiety score above $1$ and
retardation score above $2$.  This is important in settings where
patient covariates are expensive or burdensome to collect
\citep[e.g.,][]{gail1999weighing, gail2009value, 
baker2009using, huang2014characterizing}.  
%Given a function
%that quantifies the cost of measuring any subset
%of patient covariates, 
We provide an estimator of the treatment
regime that minimizes
an expected cost among all regimes
that optimize the marginal mean outcome.

% A related but different topic is the subgroup identification
% \citep{su_subgroup_2009, foster_subgroup_2011,
%   lipkovich_subgroup_2011, dusseldorp_qualitative_2014}, which aims at
% identifying a subgroup of patients in which a specific treatment shows
% enhanced treatment effect.  However, several treatments may show
% enhanced treatment effects in the same subgroup of patients, in which
% case the optimal treatment cannot be determined from the subgroup
% identification procedure.

%In Section~2, we briefly review classification-based estimation and propose an
%estimator for the optimal regime within the class of regimes
%representable as a decision list.  In Section~3, we evaluate the
%proposed estimator in a suite of simulated experiments.  In Section~4,
%we illustrate the proposed method using data from the National
%Surgical Adjuvant Breast and Bowel Project (NSABP) and
%data from a depression clinical trial. 

\section{Methodology}

\subsection{Framework}
We assume that the observed data are $\lb (X_i, A_i,
  Y_i)\rb_{i=1}^{n}$, which comprise $n$ independent identically
  distributed observations, one for each subject in an
  experimental or observational study.  Let $(X, A, Y)$ denote
a generic observation.  Then $X\in \mathbb{R}^p$ are baseline 
patient covariates; $A \in \mathcal{A} = \{ 1,\ldots, m
\}$ is the treatment assigned; and $Y\in\mathbb{R}$ is the outcome,
coded so that higher values are better.  A treatment regime,
$\pi$, is a function from $\mathbb{R}^p$ into $\mathcal{A}$, so
that under $\pi$ a patient presenting with $X=x$ is recommended treatment
$\pi(x)$.  

The value of a regime $\pi$ is the expected outcome if all patients in
the population of interest are assigned treatment according to $\pi$.
To define the value, we use the set of potential outcomes
 $\lb
  Y^*(a)\rb_{a\in\mathcal{A}}$, where $Y^*(a)$ is the
outcome that would be observed if a subject were assigned treatment
$a$.  Define $Y^*(\pi) = \sum_{a\in\mathcal{A}} Y^*(a)I\lb
  \pi(X) = a\rb$ to be the potential outcome under regime
$\pi$, and $R(\pi) = \E\lb Y^*(\pi)\rb$ 
to be the value of regime $\pi$.  An optimal
regime, say $\pi^{\mathrm{opt}}$, satisfies
$R(\pi^{\mathrm{opt}}) \geq R(\pi)$ for all $\pi$.  
Let $\Pi$ denote a class of regimes of interest. 
Classification-based estimation methods
form an estimator of $R(\pi)$, say $\widehat{R}(\pi)$, and
then estimate $\pi^{\mathrm{opt}}$ using $\widehat{\pi} =
\arg\max_{\pi\in\Pi}\widehat{R}(\pi)$.  The success of this approach
requires: (i) a high-quality estimator of $R(\pi)$; (ii) a
sufficiently rich class $\Pi$; and (iii) an efficient
algorithm for maximizing $\widehat{R}(\pi)$ over $\Pi$.  We discuss
these topics in the next three sections.

% Let $\Pi$ be the set consisting of interpretable and parsimonious
% treatment regimes to be defined later.  The optimal treatment regime
% $\pi_\text{opt}$ within this set is the one that maximize the
% expectation of $Y_i^*(\pi)$ over $\Pi$. Mathematically, we have
% \begin{equation}
% \label{eq:opt}
% \pi_\text{opt} = \argmax_{\pi \in \Pi} \E(Y_i^*(\pi)).
% \end{equation}
% In order to estimate $\pi_\text{opt}$, we need to estimate
% $\E(Y_i^*(\pi))$ using the observed data, specify the class of
% treatment regime candidates $\Pi$, and conduct the optimization.  We
% shall address these issues in subsequent subsections.

\subsection[Estimation of R(pi)]{Estimation of $R(\pi)$}
We make several standard assumptions: 
(A1) consistency: $Y = Y^*(A)$;
(A2) no unmeasured confounders: $\lb Y^*(a)
\rb_{a\in\mathcal{A}}$ are conditionally independent of
$A$ given $X$;
%written as $\lb Y^*(a) \rb_{a\in\mathcal{A}} \independent A | X$ where
%$\independent$ denotes ``independent of''; 
and (A3) positivity: there exists $\delta > 0$ so that
$\pr(A=a|X) \ge \delta$ for all $a\in\mathcal{A}$. 
Assumption (A2) is automatically
satisfied in a randomized study but is untestable in
observational studies \citep[][]{robins2000marginal}.
Under (A1)--(A3), it can be shown \citep{tsiatis2006semiparametric} that 
\begin{equation}
\label{eq:aipw1}
R(\pi) = \E \left( \sum_{a=1}^{m} \left[ \frac{I(A = a)}{\omega(X, a)} 
\left\{Y - \mu(X, a) \right\}  + \mu(X, a) \right] I\left\{
\pi(X) = a\right\} \right),
\end{equation}
where $\omega(x, a) = \pr(A=a|X=x)$ and $\mu(x, a) = \E(Y | X=x, A =
a)$.  Alternate expressions for $R(\pi)$ exist
\citep[][]{frank}; however, estimators based on~\eqref{eq:aipw1}
possess a number of desirable properties (see below).

To construct an estimator of $R(\pi)$ from~\eqref{eq:aipw1} we replace
$\omega(x, a)$ and $\mu(x, a)$ with estimated working models 
and replace the expectation with its sample analog.
If treatment is
randomly assigned independently of subject covariates, then $\omega(x,
a)$ can be estimated by $n^{-1}\sum_{i=1}^{n}I(A_i=a)$.  Otherwise, we posit a
multinomial logistic regression model of the form $\omega(x, a) =
\exp\left(u^{\T}\gamma_a\right) \big/ \big\{
  1+\sum_{j=1}^{m-1}\exp\left(u^{\T}\gamma_j\right)
\big\},\,a=1,\ldots, m-1$, where $u=u(x)$ is a known feature
vector, and $\gamma_1,\ldots, \gamma_{m-1}$ are
unknown parameters.  
 Let
$\widehat{\omega}(x, a)$ denote the maximum likelihood estimator of
$\omega(x, a)$, where $\gamma_{1},\ldots, \gamma_{m-1}$ are replaced
by maximum likelihood estimators $\widehat{\gamma}_{1},\ldots, 
\widehat{\gamma}_{m-1}$. 
%The estimator $\widehat{\omega}(x, a)$ is obtained by
%replacing $\gamma_{1},\ldots, \gamma_{m-1}$ with their maximum 
%likelihood estimators. 
We posit a generalized linear model for $\mu(x,a)$, $g
\{ \mu(x,a) \} = z^\T \beta_a$,
where $g(\cdot)$ is a known link function, $z = z(x)$ is a known
feature vector constructed from $x$, and $\beta_1,\ldots, \beta_m$ are
unknown parameters.  We use $\widehat{\mu}(x, a) =
g^{-1}(z^{\T}\widehat{\beta}_a)$ as our estimator of $\mu(x, a)$, where
$\widehat{\beta}_{1},\ldots, \widehat{\beta}_m$
are the maximum likelihood estimators of $\beta_1,\ldots,\beta_m$.

Given estimators $\widehat{\omega}(x, a)$ and $\widehat{\mu}(x, a)$, 
an estimator of $R(\pi)$ based on~\eqref{eq:aipw1} is
\begin{equation}
\label{eq:aipw2}
\widehat{R}(\pi) = \frac{1}{n}\sum_{i=1}^{n} \sum_{a=1}^{m} \left[ \frac{I(A_i = a)}
{\widehat{\omega}(X_i, a)} 
\left\{Y_i - \widehat{\mu}(X_i, a) \right\}  + \widehat{\mu}(X_i, a) \right] I\{ \pi(X_i) = a \}.
\end{equation}
For any fixed $\pi$, 
$\widehat{R}(\pi)$ is doubly robust in the sense that it is 
a consistent estimator of $R(\pi)$
if either the model for $\omega(x,a)$ or $\mu(x,a)$ is correctly
specified \citep{tsiatis2006semiparametric, zhang2012robust}.
As a direct consequence, $\widehat{R}(\pi)$ is guaranteed to 
be consistent in a randomized study, as 
$\omega(x, a)$ is known by design.  
Furthermore, if both models are correctly specified, then
$\widehat{R}(\pi)$ is semiparametric efficient; i.e., it has the
smallest asymptotic variance among the class of regular, asymptotically 
linear estimators \citep[][]{tsiatis2006semiparametric}.

% Actually, $\widehat{R}(\pi)$ is known as the doubly robust estimator of
% $R(\pi)$ \citep{tsiatis_semiparametric_2006, zhang_robust_2012}, which
% enjoys nice properties in terms of robustness and/or efficiency.  For
% any fixed treatment regime $\pi$, $\widehat{R}(\pi)$ is a consistent
% estimator of $R(\pi)$ if either the model for $\pr(A_i = a | X_i)$ or
% the model for $\E(Y_i | A_i, X_i)$ is correctly specified.  Hence
% $\widehat{R}(\pi)$ offers two chances to obtain a valid estimate of the
% true expected value of the clinical outcome.  Furthermore, 

\subsection{Regimes representable as decision lists}
\citet{gail1985testing} present an early example of a treatment
regime using data from the NSABP clinical trial.  The treatment
regime they propose is 
\begin{align*}
  &\textbf{If } \mathsf{age \leq 50\ and\ PR \leq 10} \textbf{ then } \mathsf{chemotherapy\ alone}; \\
  &\textbf{else } \mathsf{chemotherapy\ with\ tamoxifen},
\end{align*}
where age (in years) denotes the age of the patient and 
PR denotes the progesterone receptor level (in fmol).  The simple if-then
structure of the foregoing treatment regime makes it immediately interpretable.

Formally, a treatment regime, $\pi$, that is
representable as a decision list of length $L$ is described by $\lb (c_1,
a_1),\ldots, (c_L, a_L), a_0\rb$, where $c_j$ is a logical
condition that is true or false for each $x \in\mathbb{R}^p$, and
$a_j \in \mathcal{A}$ is a recommended treatment, $j=0,\ldots, L$.  
As a special case, $L = 0$ is allowed. 
The corresponding treatment regime $\{ a_0 \}$ gives
the same treatment $a_0$ to every patient.
Hereafter, let $\Pi$ denote the set of regimes that are
representable as a decision list.  Clearly, the regime proposed 
by Gail and Simon (1985) is a member of $\Pi$.

Define
$\mathcal{T}(c_j) = \lb x\in\mathbb{R}^p\,:\, c_j\,\text{is
    true for}\, x\rb$, $j=1,\ldots, L$;
$\mathcal{R}_1 = \mathcal{T}(c_1)$,
$\mathcal{R}_j = \lb \cap_{\ell < j} \mathcal{T}(c_{\ell})^{c} \rb
\bigcap \break \mathcal{T}(c_j)$, $j=2,\ldots, L$; and $\mathcal{R}_{0} = 
\bigcap_{\ell=1}^{L}
\mathcal{T}(c_{\ell})^c$, where $S^c$ is the  complement of the set $S$.  
Then a regime $\pi\in\Pi$ can be written as
$\pi(x) = \sum_{\ell=0}^{L}a_{\ell} I\left( x \in \mathcal{R}_{\ell} \right)$, which has structure 
\begin{align}
\label{eq:diagram} \nonumber
&\textbf{If } c_1 \textbf{ then } a_1; \\ \nonumber
&\textbf{else if } c_2 \textbf{ then } a_2; \\ 
&\textbf{...} \\ \nonumber
&\textbf{else if } c_{L} \textbf{ then } a_{L}; \\ \nonumber
&\textbf{else } a_0. \nonumber
\end{align}

In principle, the conditions $c_j$, and hence the sets
$\mathcal{T}(c_j)$, can be arbitrary.  
To ensure
parsimony and interpretability, we restrict $c_j$ so
that $\mathcal{T}(c_j)$ is one of the following sets:
\begin{equation}
\label{eq:cond}
\begin{aligned}
\text{[1]: }&    \{ x\in\mathbb{R}^p\,:\,x_{j_1} \leq \tau_1 \}, &
\text{[6]: }&    \{ x\in\mathbb{R}^p\,:\,x_{j_1} \leq \tau_1 \text{ or } x_{j_2} \leq \tau_2 \}, \\
\text{[2]: }&    \{x\in\mathbb{R}^p\,:\, x_{j_1} \leq \tau_1 \text{ and } x_{j_2} \leq \tau_2 \}, &
\text{[7]: }&    \{x\in\mathbb{R}^p\,:\, x_{j_1} \leq \tau_1 \text{ or } x_{j_2} > \tau_2 \}, \\
\text{[3]: }&    \{x\in\mathbb{R}^p\,:\, x_{j_1} \leq \tau_1 \text{ and } x_{j_2} > \tau_2 \}, &
\text{[8]: }&    \{x\in\mathbb{R}^p\,:\, x_{j_1} > \tau_1 \text{ or } x_{j_2} \leq \tau_2 \}, \\
\text{[4]: }&    \{x\in\mathbb{R}^p\,:\, x_{j_1} > \tau_1 \text{ and } x_{j_2} \leq \tau_2 \}, &
\text{[9]: }&    \{x\in\mathbb{R}^p\,:\, x_{j_1} > \tau_1 \text{ or } x_{j_2} > \tau_2 \}, \\
\text{[5]: }&    \{x\in\mathbb{R}^p\,:\, x_{j_1} > \tau_1 \text{ and } x_{j_2} > \tau_2 \}, &
\text{[10]: }&  \{x\in\mathbb{R}^p\,:\, x_{j_1} > \tau_1 \}, \\
\end{aligned}
\end{equation}
where $j_1 < j_2 \in \{ 1,\ldots, p \}$ are indices and
$\tau_1, \tau_2\in\mathbb{R}$ are thresholds.  We believe that the
conditions that dictate the sets in~\eqref{eq:cond}, e.g., $x_{j_1}
\leq \tau_1 \text{ and } x_{j_2} \leq \tau_2$, are more easily
interpreted than those dictated by linear thresholds, e.g., $\alpha_1
x_{j_1} + \alpha_2 x_{j_2} \leq \alpha_3$, as the former are more
commonly seen in clinical practice.

In the proposed setup,  at most two
variables are involved in any single condition.  Having a small
number of variables in each clause yields two
important properties.  First, the resulting treatment regime is
parsimonious, and the most important variables for treatment selection
are automatically identified.  Second, application of the treatment
regime allows for patient measurements to be taken in sequence so that
the treatment recommendation can be short-circuited.
For example, consider a decision list described by $\lb (c_1, a_1), (c_2, a_2), a_0 \rb$.
It is necessary to measure the variables that compose $c_1$ on all subjects,
but the variables composing $c_2$ need only be measured for those
who do not satisfy $c_1$.

\subsubsection{Uniqueness and minimal cost of a decision list}

For a decision list $\pi$ described by  
$\lb (c_1, a_1),\ldots, (c_{L}, a_{L}), a_0 \rb$,
let $\mathcal{N}_{\ell}$ denote the cost of measuring the covariates
required to check logical conditions $c_1,\ldots, c_{\ell}$.
Hereafter, for simplicity, we assume that this cost is equal to the number of
covariates needed to check $c_1,\dots, c_{\ell}$, but  it can be
extended easily to a more complex cost function reflecting risk,
burden, and availability. 
The expected cost of applying treatment rule
$\pi(x) = \sum_{\ell=0}^{L}a_{\ell}I\left( x \in \mathcal{R}_{\ell} \right)$ 
is 
$N(\pi) = \sum_{\ell=1}^{L} \mathcal{N}_{\ell} \pr\left( X \in \mathcal{R}_{\ell} \right)
 + \mathcal{N}_L \pr\left( X \in \mathcal{R}_0 \right)$, which is smaller than
$\mathcal{N}_{L} = \mathcal{N}_{L}\sum_{\ell=0}^{L}
\pr(X\in\mathcal{R}_{\ell})$, the cost of measuring all covariates in
the treatment regime. This observation reflects the benefit of the short-circuit property.

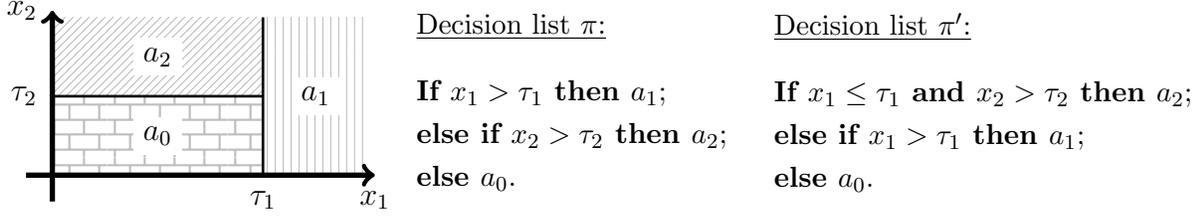
\begin{figure} 
\renewcommand{\baselinestretch}{1.1}\fontsize{11}{12}\selectfont
\begin{minipage}{0.3\linewidth}
\centering
\newcommand{\figscale}{0.35}
\newcommand{\lnth}{1pt}
%\scalefont{50}
\begin{tikzpicture}[scale=\figscale,every node/.style={font=\normalsize}]
\pattern[pattern=vertical lines,pattern color=lightgray] (0, -3) rectangle (4, 3);
\node at (2, 0) [fill=white] {$a_1$};

\pattern[pattern=north east lines, pattern color=lightgray] (0, 0) rectangle (-8, 3);
\node at (-4, 1.5) [fill=white] {$a_2$};

\pattern[pattern=bricks, pattern color=lightgray] (0, 0) rectangle (-8, -3);
\node at (-4, -1.5) [fill=white] {$a_0$};

%% Thick lines
\draw[line width=\lnth] (-8, 0) -- (0,0);
\draw[line width=\lnth] (0, -3) -- (0, 3);
\draw[line width=2*\lnth, ->] (-8, -4) -- (-8, 3.25);
\draw[line width=2*\lnth, ->] (-9, -3) -- (4.25, -3);

%% Axis labels
\node at (-8.2, 0) [anchor=east] {$\tau_2$};
\node at (0,-3.2) [anchor=north] {$\tau_1$};
\node at (-8.2, 3.25) [anchor=east] {$x_2$};
\node at (4.25, -3.2) [anchor=north] {$x_1$};
\end{tikzpicture}
\end{minipage}
\hspace{0.03\linewidth}%
\begin{minipage}{0.2\linewidth}
{\underline{Decision list $\pi$:}}
\begin{align*}  
&\textbf{If } x_1 > \tau_1 \textbf{ then } a_1; \\ 
&\textbf{else if } x_2 > \tau_2 \textbf{ then } a_2; \\ 
&\textbf{else } a_0. 
\end{align*}
\end{minipage}
\hspace{0.03\linewidth}%
\begin{minipage}{0.2\linewidth}
{\underline{Decision list $\pi'$:}}
\begin{align*}
&\textbf{If } x_1 \leq \tau_1 \textbf{ and } x_2 > \tau_2 \textbf{ then } a_2; \\
&\textbf{else if } x_1 > \tau_1 \textbf{ then } a_1; \\ 
&\textbf{else } a_0.
\end{align*}
\end{minipage}
\caption{\label{nonuniqueExample} \textbf{Left:} diagram of 
a decision list dictated by regions 
$\mathcal{R}_{1} = \lb x\in\mathbb{R}^2\,:\, x_1 > \tau_1\rb$, 
$\mathcal{R}_2 = \lb x \in\mathbb{R}^2\,:x_1 \leq \tau_1,\,x_2 > \tau_2\rb$, and
$\mathcal{R}_{0} = \lb x\in\mathbb{R}^2\,:\, x_1 \leq \tau_1,\,x_2 \leq \tau_2\rb$, and
treatment recommendations $a_1$, $a_2$, and $a_0$. \textbf{Middle:} 
representation of the decision list that requires only
$x_1$ in the first clause.  \textbf{Right:} alternative representation
of the same decision list that requires both $x_1$ and $x_2$ in the first
clause.}  
\end{figure}

A decision list $\pi$ described by
$\lb (c_1, a_1),\ldots, (c_{L}, a_{L}), a_0 \rb$ need not be unique in
that there may exist an alternative decision list $\pi'$ described by
$\lb (c_1', a_1'),\ldots, (c_{L'}', a_{L'}'), a_0' \rb$ such that
$\pi(x) = \pi'(x)$ for all $x$ but $L \ne L'$, or $L=L'$ but
$c_j \ne c_j'$ or $a_j \ne a_j'$ for some $j \in \{ 1,\ldots, L \}$.
This is potentially important because the expected costs $N(\pi)$ and
$N(\pi')$ might differ substantially.  Figure~\ref{nonuniqueExample}
shows two representations, $\pi$ and $\pi'$, of the same decision list
both with $L=L'=2$ but with different clauses.  The cost of the
decision list in the middle panel, $\pi$, is
$N(\pi) = \mathcal{N}_{1} \pr(X_1 > \tau_1) + \mathcal{N}_2 \pr(X_1
\le \tau_1)$,
whereas the cost of the decision list in the right panel, $\pi'$, is
$N(\pi') = \mathcal{N}_{2} \ge N(\pi)$ with strict inequality if
$\mathcal{N}_2 > \mathcal{N}_1$ and $\pr(X_1 > \tau_1) > 0$.  
Thus, $\pi$ is preferred to $\pi'$ in settings where 
$X_2$ is a biomarker that is expensive, burdenome,
or potentially harmful to collect
\citep[e.g.,][and references therein]{ huang2014characterizing}.  
%Thus, if $X_2$ is a biomarker that is expensive to collect, we are
%able to avoid such a burden for the group of patients satisfying
%$X_1 > \tau_1$ by using decision list $\pi$ instead of $\pi'$.

Therefore, among all decision lists achieving the value $R(\pi^{\mathrm{opt}})$,
where $\pi^{\mathrm{opt}}$ is an optimal regime as defined previously,
we seek to estimate the one that minimizes the cost.
Defining $\mathcal{L}_{r}$ to be the
level set $\big\{ \pi\in\Pi\,:\, R(\pi) = r \big\}$, 
then the goal is to estimate a regime in the set
$\arg\min_{\pi \in \mathcal{L}\lb R (\pi^{\mathrm{opt}})\rb}N(\pi)$.  
Define $\widehat{\mathcal{L}}(r) = \big\{ \pi\in\Pi\,:\,\widehat{R}(\pi) = r\big\}$.  
Let $\widetilde{\pi}$ be
an estimator of an element in the set $\arg\max_{\pi\in\Pi}\widehat{R}(\pi)$. 
In the following we provide
an algorithm that ensures our estimator, $\widehat{\pi}$, belongs to the set
$\arg\min_{\pi \in \widehat{\mathcal{L}}\lb \widehat{R}(\widetilde{\pi})\rb}
\wh{N}(\pi)$, where $\wh{N}(\pi)$ is defined by replacing the probabilities in $N(\pi)$ with sample proportions.

% \end{figure}
% Consider the decision list described by $\lb (c_1,
%   a_1),\ldots, (c_L, a_L), a_0\rb$, if for some $\ell$ the
% logical conditions $\bigcap_{j=1}^{\ell-1}c_j^c$ and $c_{j+1}$ are
% disjoint then $\mathcal{R}_{j+1} = \emptyset $ and thus the decision
% list is not well-defined as $\mathcal{R}_{j+1}$ could be replaced by
% any other set that is disjoint from $\bigcap_{j=1}^{\ell-1}c_j^c$
% without altering the treatment recommendations of the decision list.
% The estimation procedure described in the next section uses a greedy
% stepwise optimization algorithm that automatically avoids vacuous
% logical conditions.  

\subsection{Computation}
Estimation proceeds in two steps: (i) approximate
an element $\widetilde{\pi} \in \arg\max_{\pi\in\Pi}\widehat{R}(\pi)$, 
where $\widehat{R}(\pi)$ is constructed using~\eqref{eq:aipw2}; 
and (ii) find an element $\widehat{\pi} \in
\arg\min_{\pi \in \widehat{\mathcal{L}}\lb
    \widehat{R}(\widetilde{\pi})\rb} \wh{N}(\pi)$.

\subsubsection[Approximation of arg max R(pi)]{Approximation of $\arg\max_{\pi\in\Pi}\widehat{R}(\pi)$}

Maximizing $\widehat{R}(\pi)$ over $\pi\in\Pi$ is computationally
burdensome in problems with more than a handful of covariates because of the
indicator functions in~\eqref{eq:aipw2} and the discreteness of the
decision list.  However, the tree structure of decision lists suggests
a greedy algorithm in the spirit of classification and regression
trees (CART, \citealp{breiman1984classification}).  
Assume that for the $j$th covariate, there is a candidate set of
finitely many possible cutoff values $\mathcal{X}_j$.  These cutoffs
might be dictated by clinical guidelines, e.g., if the covariate is a
comorbid condition then the thresholds might reflect low, moderate,
and high levels of impairment; alternatively, these cutoffs could be
chosen to equal empirical or theoretical percentiles of that
covariate.  There is no restriction imposed on these cutoffs.  Let
$\mathcal{C}$ denote the set of all conditions that induce regions of the
form in~\eqref{eq:cond} with the cutoffs $\tau_{j_k} \in
\mathcal{X}_{j_k}$ for $k=1,2$, $j_k \in \{ 1,\ldots, p \}$.

Before giving the
details of the algorithm, we provide a conceptual overview.
The algorithm first uses exhaustive search to find a decision list 
with exactly one clause, of the form $\pi = \lb (c_1, a_1), a_1' \rb$, 
which maximizes $\widehat{R}(\pi)$.  Let $\lb (\widetilde{c}_1, 
\widetilde{a}_1), \widetilde{a}_1'\rb$ denote
this decision list.  The algorithm then uses exhaustive
search to find the decision list that maximizes $\widehat{R}(\pi)$
over decision lists with exactly two clauses, the first of
which must be either $(\widetilde{c}_1, \widetilde{a}_1)$ or $(\widetilde{c}_1', \widetilde{a}_1')$,
where $\widetilde{c}_1'$ is the negation of $\widetilde{c}_1$ 
such that $\mathcal{T}(\widetilde{c}_1') = \mathcal{T}(\widetilde{c}_1)^c$;
e.g., if $\widetilde{c}_1$ has the form $x_{j_1} \leq \tau_1$ and $x_{j_2} \leq \tau_2$,
then $\widetilde{c}_1'$ would be $x_{j_1} > \tau_1$ or $x_{j_2} > \tau_2$.
Although the decision lists $\lb (\widetilde{c}_1, \widetilde{a}_1), \widetilde{a}_1'\rb$ and
$\lb (\widetilde{c}_1', \widetilde{a}_1'), \widetilde{a}_1\rb$ yield identical treatment recommendations and have the same value,
their first clauses are distinct, and may lead to substantially different final decision lists.
Hence it is necessary to consider both possibilities for the first clause.
The algorithm proceeds
recursively by adding one clause at a time until some stopping
criterion (described below) is met. 

Hereafter, for a decision list $\pi$ described by $\lb (c_1,
a_1),\ldots, (c_L, a_L), a_0\rb$ for some $L\ge 0$, write\break
$\widehat{R}\left[ \lb (c_1, a_1),\ldots, (c_L, a_L), a_0\rb \right]$
to denote $\widehat{R}(\pi)$; e.g., for $L=0$, $\widehat{R}\left[\lb
  a_0\rb\right]$ is the estimated value of the regime that assigns
treatment $a_0$ to all patients.  For any decision list with a vacuous
condition, e.g., $\lb \cap_{\ell < j} \mathcal{T}(c_{\ell})^{c} \rb
\bigcap \mathcal{T}(c_j) = \emptyset$ for some $j$, define
$\widehat{R}\left[ \lb (c_1, a_1),\ldots, (c_L, a_L), a_0\rb \right] =
-\infty$.  Let $z_{\rho}$ be the $100 \rho$ percentile of the standard
normal distribution.  Let $\Pi_\text{temp}$ denote the set of regimes
to which additional clauses can be added, and let $\Pi_\text{final}$
denote the set of regimes that have met one of the stopping criteria.
The algorithm is as follows, and an illustrative example with a
step-by-step run of the algorithm is given in the Web
Appendix.
\vspace{-0.25in}
\begin{enumerate}[leftmargin=0in]
\item[] \textbf{Step~1.} 
Choose a maximum list length $L_{\max}$ and a critical level $\alpha \in (0, 1)$.
Compute $ \wt{a}_{0} = \argmax_{a_0\in\mathcal{A}} \wh{R}\left[ \lb a_0 \rb \right] $.
Set $\Pi_\text{temp} = \emptyset$ and $\Pi_\text{final} = \emptyset$.

\item[] \textbf{Step~2.} 
Compute $ 
  (\wt{c}_{1}, \wt{a}_{1}, \wt{a}_{1}') = 
  \argmax_{(c_{1}, a_{1}, a_{1}') \in \mathcal{C} \times \mathcal{A} \times \mathcal{A}}
  \wh{R}\left[ \lb (c_{1}, a_{1}), a_{1}' \rb \right] 
$ and $ 
  \wh{\Delta}_1 = \break \wh{R}\left[ \lb (\wt{c}_{1}, \wt{a}_{1}), \wt{a}_{1}' \rb \right] - \wh{R}\left[ \lb \wt{a}_0 \rb \right]
$.
If $\wh{\Delta}_1 < z_{1-\alpha} \big\{ \wh{\var} \big(\wh{\Delta}_1\big) \big\}^{1/2}$ then
let $ \pi = \lb \wt{a}_0 \rb $, set $\Pi_\text{final} = \{ \pi \}$, and go to Step~5;
otherwise let $ \pi = \lb (\wt{c}_{1}, \wt{a}_{1}), \wt{a}_{1}' \rb $,
$ \pi' = \lb (\wt{c}_{1}', \wt{a}_{1}'), \wt{a}_{1} \rb $,
set $\Pi_\text{temp} = \{ \pi, \pi' \}$, and proceed to Step~3,
where $\wt{c}_{1}'$ is the negation of $\wt{c}_{1}$.

\item[] \textbf{Step~3.} 
Pick an element $\wb{\pi} \in \Pi_\text{temp}$, say
$ \wb{\pi} = \lb (\wb{c}_1, \wb{a}_1), \dots, (\wb{c}_{j-1}, \wb{a}_{j-1}), \wb{a}_{j-1}' \rb $, 
where $j-1$ is the length of $\wb{\pi}$.
Remove $\wb{\pi}$ from $\Pi_\text{temp}$.
With the clauses $(\wb{c}_1, \wb{a}_1), \dots, (\wb{c}_{j-1}, \wb{a}_{j-1})$ held fixed, 
compute $ 
  (\wt{c}_{j}, \wt{a}_{j}, \wt{a}_{j}') = 
    \argmax_{(c_{j}, a_{j}, a_{j}') \in \mathcal{C} \times \mathcal{A} \times \mathcal{A}} 
    \wh{R}\left[ \lb (\wb{c}_1, \wb{a}_1),\dots, (\wb{c}_{j-1}, \wb{a}_{j-1}), (c_j, a_j), a_j' \rb \right]
$ and $
  \wh{\Delta}_j = 
    \wh{R}\left[ \lb (\wb{c}_1, \wb{a}_1),\dots, (\wb{c}_{j-1}, \wb{a}_{j-1}), (\wt{c}_j, \wt{a}_j), \wt{a}_j' \rb \right]
    - \wh{R}\left[ \lb (\wb{c}_1, \wb{a}_1),\dots, (\wb{c}_{j-1}, \wb{a}_{j-1}), \wb{a}_{j-1}' \rb \right]
$.
If $\wh{\Delta}_j < z_{1-\alpha} \big\{ \wh{\var}\big(\wh{\Delta}_j\big) \big\}^{1/2}$, then
let $\pi = \lb (\wb{c}_1, \wb{a}_1), \dots, (\wb{c}_{j-1}, \wb{a}_{j-1}), \wb{a}_{j-1}' \rb$,
and set $\Pi_\text{final} = \Pi_\text{final} \bigcup \{ \pi \}$;
otherwise if $j = L_\mathrm{max}$,
let $\pi = \lb (\wb{c}_1, \wb{a}_1), \dots, (\wb{c}_{j-1}, \wb{a}_{j-1}), (\wt{c}_j, \wt{a}_j), \wt{a}_j' \rb$,
and set $\Pi_\text{final} = \Pi_\text{final} \bigcup \{ \pi \}$;
otherwise set $\Pi_\text{temp} = \Pi_\text{temp} \bigcup \{ \pi, \pi' \}$,
where $\wt{c}_j'$ is the negation of $\wt{c}_j$, $\pi = \lb (\wb{c}_1, \wb{a}_1), \dots, (\wb{c}_{j-1}, \wb{a}_{j-1}), (\wt{c}_j, \wt{a}_j), \wt{a}_j' \rb$, and
$\pi' = \lb (\wb{c}_1, \wb{a}_1), \dots, (\wb{c}_{j-1}, \wb{a}_{j-1}), (\wt{c}_j', \wt{a}_j'), \wt{a}_j \rb$.

\item[] \textbf{Step~4.} 
Repeat Step~3 until $\Pi_\text{temp}$ becomes empty.

\item[] \textbf{Step~5.} 
Compute $\wt{\pi} = \argmax_{\pi \in \Pi_\text{final}} \wh{R}(\pi)$. Then $\wt{\pi}$ is the estimated optimal decision list.
\end{enumerate}

The above description is simplified to illustrate the main ideas.
The actual implementation of this algorithm avoids exhaustive searches by
pruning the search space $\mathcal{C} \times \mathcal{A} \times \mathcal{A}$.
It also avoids explicit construction of $\Pi_\text{temp}$ and $\Pi_\text{final}$.
Complete implementation details are provided in the Web Appendix.  
In the algorithm, the decision list stops growing 
if either the estimated increment in the value, $\wh{\Delta}_j$, is not sufficiently large compared to 
an estimate of its variation, $\big\{ \wh{\var}\big(\wh{\Delta}_j\big) \big\}^{1/2}$,
or if it reaches the pre-specified maximal length $L_\text{max}$.
We estimate $\var(\widehat{\Delta}_j)$ using large sample theory; the expression is given
in the Web Appendix. This variance estimator
is a crude approximation, as it ignores uncertainty
introduced by the estimation of the decision lists; however,
it can be computed quickly, and in simulated experiments it
appears sufficient for use in a stopping criterion.  The
significance level $\alpha$ is a user-chosen tuning parameter.
In our simulation experiments, we chose $\alpha=0.05$; 
results were not sensitive to this choice (see Web Appendix). 
To avoid lengthy lists, we set $L_\text{max}=10$.
Nevertheless, in our simulations and applications the estimated lists
never reach this limit.
Finally, it may be desirable in practice to restrict the set of 
candidate clauses so that, for each $j$, the number of subjects in
$\widehat{\mathcal{R}}_{j} = \lb \cap_{\ell < j}
\mathcal{T}(\widehat{c}_{\ell})^{c}\rb
\bigcap \mathcal{T}(\widehat{c}_{j})$ exceeds 
some minimal threshold.  This can be readily incorporated
into the above algorithm by simply discarding candidate
clauses that induce partitions that contain an insufficient
number of observations.  

The time complexity of the proposed algorithm is $O\left[
2^{L_{\max}} m p^2
\lb n + \left(\max_{j}\#\mathcal{X}_{j}\right)^2\rb\right]$ (see Web Appendix),
where $\# \mathcal{X}_{j}$ is the number of cutoff values in
$\mathcal{X}_{j}$. Because
$2^{L_{\max}}$ and $m$ are constants that are typically small relative to
$p^2\lb n + \left(\max_{j}\#\mathcal{X}_{j}\right)^2\rb$, 
the time
complexity is essentially $O(n p^2)$ provided that
$\max_{j}\#\mathcal{X}_{j}$ is either fixed or diverges more slowly than
$n^{1/2}$.  Hence, the time complexity is the same as a single least
squares fit, indicating that the proposed algorithm runs very fast and
scales well in both dimension $p$ and sample size $n$.

\subsubsection[Finding an element of arg min N(pi)]{Finding an element of 
$\arg\min_{\pi\in\widehat{\mathcal{L}}\lb \widehat{R}(\widetilde{\pi})\rb}
\wh{N}(\pi)$}
To find an element within the set
$\arg\min_{\pi\in\widehat{\mathcal{L}}\lb \widehat{R}(\widetilde{\pi})\rb} \wh{N}(\pi)$,
we enumerate all regimes in
$\mathcal{L}\big\{ \widehat{R}(\widetilde{\pi}) \big\}$ 
with length no larger
than $L_{\max}$ and select among them the list with the minimal cost. 
The enumeration algorithm is recursive and requires
a substantial amount of bookkeeping; therefore, we describe the basic
idea here and defer implementation details to the Web Appendix. 
Suppose $\widetilde{\pi}$ is described by $\lb
  (\widetilde{c}_{1}, \widetilde{a}_{1}), \ldots, (
  \widetilde{c}_{L}, \widetilde{a}_{L}), 
  \widetilde{a}_{0} \rb$.  Call a condition of
the form $x_{j} \le \tau_{j}$ an atom. There exist $K \leq 2L$ atoms,
say $d_1,\ldots, d_K$, such that each clause $\widetilde{c}_{\ell}$,
$\ell=1, \ldots, L$, can be expressed using the union, intersection,
and/or negation of at most two of these atoms.  The algorithm proceeds
by generating all lists with clauses representable using the foregoing
combinations of at most two atoms.  To reduce computation time, we use
a branch-and-bound scheme \citep[][]{brusco2006branch} that avoids
constructing lists with vacuous conditions or those that are provably
worse than an upper bound on
$\min_{\pi\in\widehat{\mathcal{L}}\lb \widehat{R}(\widetilde{\pi})\rb}N(\pi)$.
In the simulation experiments in the next section, the average runtime
for the enumeration algorithm was less than one second running on a
single core of a 2.3GHz AMD Opteron\texttrademark{} processor and 
1GB of DDR3 RAM.

\section{Simulation Experiments}

We use a series of simulated experiments to examine the finite sample
performance of the proposed method.  The average value
$\mathbb{E}\lb R(\widehat{\pi})\rb$ and the average
cost $\mathbb{E}\lb N(\widehat{\pi})\rb$ 
are the primary performance
measures.  We consider generative models with 
(i) binary and continuous outcomes; (ii) binary and trinary
treatments; (iii) correctly and incorrectly specified models; and (iv)
low- and high-dimensional covariates.  The class of data-generating
models that we consider is as follows.  Covariates are drawn from a
$p$-dimensional Gaussian distribution with mean zero and
autoregressive covariance matrix such that $\mathrm{cov}(X_k, X_\ell) =
4(1/5)^{|k-\ell|}$, and the treatments are sampled uniformly so that
$P(A=a|X=x) = 1/m$ for all $x \in \mathbb{R}^p$ and $a \in \mathcal{A}$.  Let $\phi(x,
a)$ be a real-valued function of $x$ and $a$; given $X=x$ and $A=a$,
continuous outcomes are normally distributed with mean $2 + x_{1} +
x_{3} + x_{5} + x_{7} + \phi(x, a)$ and variance $1$, whereas binary
outcomes follows a Bernoulli distribution with success probability $\expit\lb 2
  + x_{1} + x_{3} + x_{5} + x_{7} + \phi(x, a)\rb$, where
$\expit(u) = \exp(u)/\{1+\exp(u)\}$.  
Table~\ref{contrastTable} lists the expressions of $\phi$ used in our
generative models and the number of treatments, $m$, in $\mathcal{A}$.
Under these outcome models, the optimal regime is
$\pi^{\mathrm{opt}}(x) = \arg\max_{a} \phi(x, a)$.

For comparison, we estimate $\pi^{\mathrm{opt}}$ by
parametric $Q$-learning, nonparametric $Q$-learning, outcome weighted
learning (OWL, \citealp{zhao2012estimating}) and modified covariate
approach (MCA, \citealp{tian2014simple}).  For parametric
$Q$-learning, we use linear regression when $Y$ is continuous and
logistic regression when $Y$ is binary.  The linear component in the
regression model has the form
$\sum_{a=1}^{m} I(A = a) (1, X^\T) \beta_a$, where
$\beta_{1},\dots,\beta_{m}$ are unknown coefficient vectors.  A LASSO
penalty \citep[][]{tibshirani1996regression} is used to reduce
overfitting; the amount of penalization is chosen by minimizing 10-fold
cross-validated prediction error.  For nonparametric $Q$-learning, we
use support vector regression when $Y$ is continuous and support
vector machines when $Y$ is binary
\citep[][]{zhao2011reinforcement},
both are implemented using a Gaussian kernel.
Tuning parameters for non-parametric $Q$-learning are selected by minimizing
10-fold cross-validated prediction error.  For OWL, both linear and
Gaussian kernels are used and we follow the same tuning strategy as in
\citet{zhao2012estimating}.  For MCA, we incorporate the efficiency
augmentation term described in \citet{tian2014simple}.  Both OWL and
MCA are limited to two treatment options. 

To implement our method, the mean model, $\mu(x, a)$,
in~\eqref{eq:aipw1}, is estimated as in parametric
$Q$-learning.  The propensity score $\omega(x, a)$ is estimated by
$n^{-1}\sum_{i=1}^{n} I(A_i = a)$.  All the comparison methods result
in treatment regimes 
that are more difficult to interpret than a
decision list; thus, our intent is to show that decision lists are
competitive in terms of the achieved value of the
estimated regime, $\mathbb{E}\{ R(\widehat{\pi}) \}$, while being
significantly more interpretable and less costly.

Results in Table~\ref{tbl:results} are based on the
average over 1000 Monte Carlo replications with
data sets of size $n=500$ if $m=2$ and $Y$ is continuous; $n=750$ if
$m=3$ and $Y$ is continuous; $n=1000$ if $m=2$ and $Y$ is binary; and
$n=1500$ if $m=3$ and $Y$ is binary.  
The 
value $R(\wh{\pi})$ and cost $N(\wh{\pi})$ 
were computed using an independent test set of size $10^6$.% subjects.
%The values and costs from 1000 replications were averaged.
%Because OWL and MCA are not applicable when there are three treatment options,
%some rows of Table~\ref{tbl:results} contain missing values.

Table~\ref{tbl:results} shows that the decision list is
competitive in terms of the value
obtained across the entire suite of simulation experiments.  
If $\pi^{\mathrm{opt}}$ can be
represented as a decision list, the proposed method produces
the best value.  However, even in settings in which the 
optimal regime is not a decision list, the estimated 
decision list appears to perform well.  Recall that the
proposed algorithm attempts to find the best approximation
of the optimal
regime within the class of regimes that are representable as a decision list.
Figure~\ref{fig:misspecified} 
shows the average estimated decision list in 
misspecified settings \RomanNum{2} and \RomanNum{3} with continuous outcome
and $p=10$.    In these settings, the estimated decision list 
provides a reasonable approximation of the true optimal regime.
% Otherwise, the proposed method attempts to approximate
% $\pi^{\mathrm{opt}}$ using decision lists.  The computation algorithm
% in Section~2.4, which doesn't require $\pi^{\mathrm{opt}}$ be
% representable as a decision list, still produces reasonable regimes.
% For example, Figure~\ref{fig:misspecified} shows the estimated regimes
% under setting \RomanNum{2} and \RomanNum{3} with continuous outcome
% and $p=10$, averaged over 1000 replications.
%Although
%$\pi^{\mathrm{opt}}$ in these settings is not in the form of a
%decision list, the estimated regime, which is a decision list,
%captures the treatment decision boundary quite well and provides
%correct recommendation for a large proportion of the patients.  
In addition, the cost of the decision list 
is notably smaller than the cost of the parametric $Q$-learning
estimator or the MCA estimator.  Nonparametric $Q$-learning
OWL always use all covariates, so
their costs are always equal to $p$.  

%Therefore, in
%the simulated examples considered, the decision list provides good
%performance in terms of value while being interpretable and 

In the Web Appendix, we derive point estimates and
prediction intervals for $R(\wh{\pi})$.  We also present simulation
results to illustrate the accuracy of variable selection for the
decision list.

\begin{table}
\renewcommand{\baselinestretch}{1.1}\fontsize{11}{12}\selectfont
\caption{The second column gives the number of treatment options $m$.
  The third column gives the set of $\phi$ functions used in the outcome models. 
  The fourth column specifies the form of the optimal regime 
  $\pi^{\mathrm{opt}}(x) = \arg\max_{a}\phi(x,a)$
  where: ``linear'' indicates that $\pi^{\mathrm{opt}}(x) = 
  \arg\max_{a}\{(1,x^{\T})\beta_{a}\}$ for some coefficient vectors $\beta_{a}\in\mathbb{R}^{p+1}$,
  $a\in\mathcal{A}$; 
  ``decision list'' indicates that $\pi^{\mathrm{opt}}$ is representable
  as a decision list; 
  and ``nonlinear'' indicates that $\pi^{\mathrm{opt}}(x)$
  is neither linear nor representable as a decision list.}
\label{contrastTable}
\begin{tabular*}{\columnwidth}{c @{\extracolsep{\fill}} c @{\extracolsep{\fill}} p{8.5cm} @{\extracolsep{\fill}} l}
  \toprule
Setting & $m$ & \cc{Expression of $\phi$} & \cc{Form of $\pi^\mathrm{opt}$} \\  
  \midrule
\RomanNum{1} & $2$ & 
   $\phi_1(x, a) = I(a = 2) \{ 3 I(x_{1} \leq 1, x_{2} > -0.6) - 1 \}$ & decision list \\
\RomanNum{2} & $2$ & 
   $\phi_2(x, a) = I(a = 2) (x_{1} + x_{2} - 1)$ & linear \\
\RomanNum{3} & $2$ & 
   $\phi_3(x, a) = I(a = 2) \arctan (\exp(1 + x_{1}) - 3 x_{2} - 5)$ & nonlinear \\
\RomanNum{4} & $2$ & 
   $\phi_4(x, a) = I(a = 2) (x_{1} - x_{2} + x_{3} - x_{4})$ & linear \\ 
\RomanNum{5} & $3$ & 
   $\begin{matrix}\phi_5(x, a) = I(a = 2) \{4 I(x_{1} > 1) - 2\} \phantom{12345678} \\  {} + I(a = 3) I(x_{1} \leq 1) \{ 2 I(x_{2} \leq -0.3) - 1 \} \end{matrix}$ & decision list \\ 
\RomanNum{6} & $3$ & 
   $\phi_6(x, a) = I(a = 2) (2 x_{1}) + I(a = 3) (-x_{1} x_{2})$ & nonlinear \\
\RomanNum{7} & $3$ & 
   $\phi_7(x, a) = I(a = 2) (x_{1} - x_{2}) + I(a = 3) (x_{3} - x_{4})$ & linear \\ 
  \bottomrule
\end{tabular*}
\end{table}

\begin{table}
\renewcommand{\baselinestretch}{1.1}\fontsize{11}{12}\selectfont
\caption{The average value and the average cost of estimated regimes in simulated experiments.
In the header, $p$ is the dimension of patient covariates;
DL refers to the proposed method using decision list;
$Q_1$ refers to parametric $Q$-learning;
$Q_2$ refers to nonparametric $Q$-learning;
$\text{OWL}_1$ and $\text{OWL}_2$ refer to outcome weighted learning with linear kernel and Gaussian kernel, respectively;
MCA refers to modified covariate approach with efficiency augmentation.
OWL and MCA are not applicable under Setting \RomanNum{5}, \RomanNum{6} and \RomanNum{7}.}
\label{tbl:results}
\begin{tabular*}{\columnwidth}{*{10}{c @{\extracolsep{\fill}}} c}
  \toprule
\multirow{2}{*}{$p$} & \multirow{2}{*}{Setting}
  & \multicolumn{6}{c}{Value} & \multicolumn{3}{c}{Cost} \\
  \cmidrule{3-8} \cmidrule{9-11}
  &
  & DL & $Q_1$ & $Q_2$ & $\text{OWL}_1$ & $\text{OWL}_2$ & MCA
  & DL & $Q_1$ & MCA \\ 
  \midrule
\multicolumn{5}{l}{\emph{Continuous response}} \\
  \midrule
  \multirow{7}{*}{$10$} 
  & \RomanNum{1} & 2.78 & 2.53 & 2.53 & 2.33 & 2.29 & 2.54 & 1.64 & \phantom{0}9.0 & \phantom{0}5.1 \\ 
  & \RomanNum{2} & 2.70 & 2.80 & 2.79 & 2.61 & 2.54 & 2.80 & 1.64 & \phantom{0}9.0 & \phantom{0}5.1 \\ 
  & \RomanNum{3} & 2.59 & 2.54 & 2.53 & 2.29 & 2.24 & 2.55 & 1.68 & \phantom{0}9.1 & \phantom{0}4.9 \\ 
  & \RomanNum{4} & 2.89 & 3.37 & 3.35 & 3.16 & 3.09 & 3.37 & 2.50 & \phantom{0}9.5 & \phantom{0}7.4 \\ 
  & \RomanNum{5} & 2.90 & 2.67 & 2.59 & $-$ & $-$ & $-$ & 1.90 & \phantom{0}9.5 & $-$ \\ 
  & \RomanNum{6} & 3.98 & 3.46 & 3.95 & $-$ & $-$ & $-$ & 1.61 & \phantom{0}9.2 & $-$ \\ 
  & \RomanNum{7} & 3.22 & 3.75 & 3.73 & $-$ & $-$ & $-$ & 2.56 & \phantom{0}9.7 & $-$ \\ 
  \midrule
  \multirow{7}{*}{$50$} 
  & \RomanNum{1} & 2.76 & 2.51 & 2.36 & 2.21 & 2.19 & 2.53 & 1.80 & 21.3 & \phantom{0}9.2 \\ 
  & \RomanNum{2} & 2.70 & 2.79 & 2.73 & 2.26 & 2.27 & 2.79 & 1.64 & 21.4 & \phantom{0}9.3 \\ 
  & \RomanNum{3} & 2.59 & 2.52 & 2.35 & 2.16 & 2.12 & 2.54 & 1.71 & 23.1 & \phantom{0}9.0 \\ 
  & \RomanNum{4} & 2.89 & 3.36 & 3.27 & 2.76 & 2.70 & 3.36 & 2.53 & 25.4 & 14.9 \\ 
  & \RomanNum{5} & 2.87 & 2.63 & 2.33 & $-$ & $-$ & $-$ & 2.14 & 28.5 & $-$ \\ 
  & \RomanNum{6} & 3.95 & 3.43 & 3.47 & $-$ & $-$ & $-$ & 1.69 & 26.6 & $-$ \\ 
  & \RomanNum{7} & 3.21 & 3.74 & 3.61 & $-$ & $-$ & $-$ & 2.55 & 30.8 & $-$ \\ 
  \midrule
\multicolumn{5}{l}{\emph{Binary response}} \\
  \midrule
  \multirow{7}{*}{$10$} 
  & \RomanNum{1} & 0.77 & 0.74 & 0.69 & 0.73 & 0.73 & 0.74 & 1.94 & \phantom{0}8.9 & \phantom{0}4.1 \\ 
  & \RomanNum{2} & 0.71 & 0.72 & 0.60 & 0.71 & 0.71 & 0.72 & 1.69 & \phantom{0}9.2 & \phantom{0}5.3 \\ 
  & \RomanNum{3} & 0.73 & 0.73 & 0.68 & 0.72 & 0.72 & 0.73 & 2.10 & \phantom{0}9.2 & \phantom{0}4.7 \\ 
  & \RomanNum{4} & 0.71 & 0.76 & 0.66 & 0.75 & 0.74 & 0.75 & 2.40 & \phantom{0}9.6 & \phantom{0}8.4 \\ 
  & \RomanNum{5} & 0.75 & 0.73 & 0.62 & $-$ & $-$ & $-$ & 2.52 & \phantom{0}9.6 & $-$ \\ 
  & \RomanNum{6} & 0.79 & 0.75 & 0.64 & $-$ & $-$ & $-$ & 2.09 & \phantom{0}9.5 & $-$ \\ 
  & \RomanNum{7} & 0.77 & 0.81 & 0.69 & $-$ & $-$ & $-$ & 2.83 & \phantom{0}9.9 & $-$ \\ 
  \midrule
  \multirow{7}{*}{$50$} 
  & \RomanNum{1} & 0.76 & 0.73 & 0.69 & 0.71 & 0.70 & 0.73 & 2.64 & 21.9 & \phantom{0}8.3 \\ 
  & \RomanNum{2} & 0.71 & 0.72 & 0.60 & 0.70 & 0.69 & 0.71 & 1.87 & 26.2 & \phantom{0}6.4 \\ 
  & \RomanNum{3} & 0.73 & 0.72 & 0.67 & 0.70 & 0.69 & 0.72 & 2.53 & 25.0 & \phantom{0}7.3 \\ 
  & \RomanNum{4} & 0.71 & 0.76 & 0.66 & 0.73 & 0.72 & 0.74 & 2.55 & 31.0 & 13.8 \\ 
  & \RomanNum{5} & 0.74 & 0.72 & 0.61 & $-$ & $-$ & $-$ & 3.15 & 30.4 & $-$ \\ 
  & \RomanNum{6} & 0.78 & 0.75 & 0.63 & $-$ & $-$ & $-$ & 2.41 & 29.6 & $-$ \\ 
  & \RomanNum{7} & 0.76 & 0.81 & 0.68 & $-$ & $-$ & $-$ & 2.97 & 35.7 & $-$ \\ 
  \bottomrule
\end{tabular*}
\end{table}

\begin{figure}
\renewcommand{\baselinestretch}{1.1}\fontsize{11}{12}\selectfont
\centering
\includegraphics[scale = 0.7]{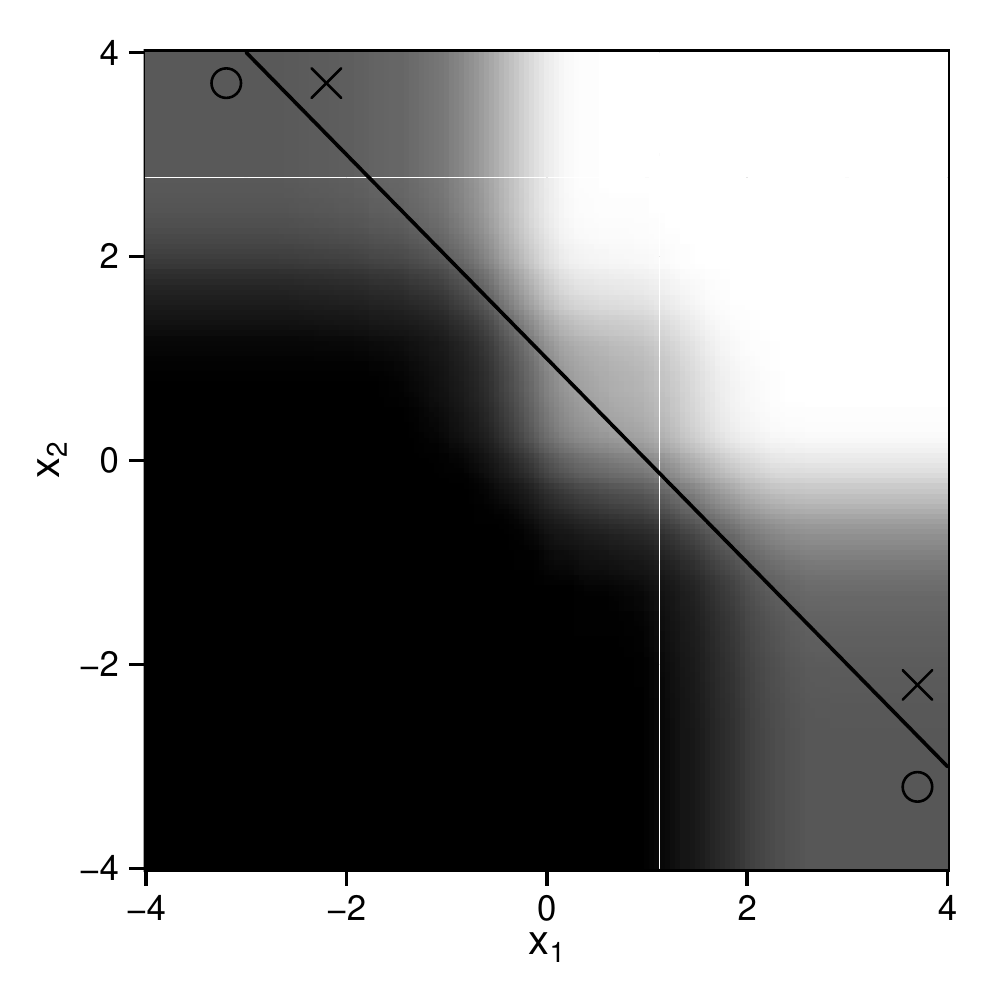}\qquad
\includegraphics[scale = 0.7]{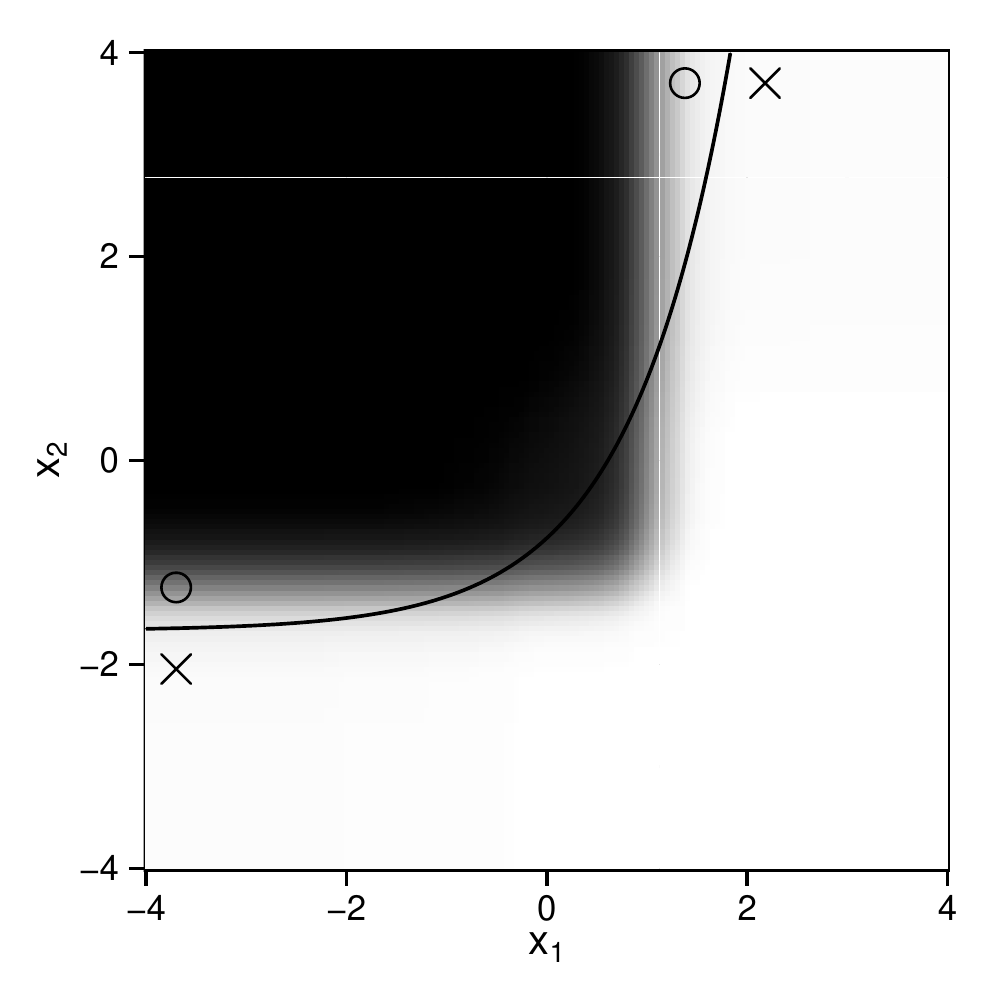}
\caption{\textbf{Left: }average estimated regimes under setting~\RomanNum{2}. 
\textbf{Right: }average estimated regimes under setting~\RomanNum{3}.
In both settings $\pi^{\mathrm{opt}}$ cannot be represented as decision list.
The solid line is the treatment decision boundary under $\pi^{\mathrm{opt}}$.
The region where treatment~1 is better than treatment~2 is marked by circles,
while the region where treatment~2 is better than treatment~1 is marked by crosses.
For every point $(x_1, x_2)^\T$, we compute the proportion of 1000 replications that the estimated regime
recommends treatment~1 to a patient with covariate $(x_1, x_2, 0, \dots, 0) \in \mathbb{R}^{10}$.
The larger the proportion, the darker the shade.
}
\label{fig:misspecified}
\end{figure}

\section{Applications}

% In this section we apply the proposed method to two data sets from two
% randomized clinical trials respectively, one dealing with the breast
% cancer and the other involving the chronic depression.

\subsection{Breast Cancer Data}
%As discussed previously,
\citet{gail1985testing} compared the
treatment effects of chemotherapy alone and chemotherapy with
tamoxifen using data collected from the NSABP trial.  Their regime
recommended chemotherapy alone to patients with $\text{age} \leq 50$
and $\text{PR} \leq 10$ and chemotherapy plus tamoxifen to all others.
Because the variables involved in the treatment regime constructed by
Gail and Simon were chosen using clinical judgment, it is of interest
to see what regime emerges from a more data-driven procedure.  Thus,
we use the proposed method to estimate an optimal treatment regime in the form of a decision list using data
from the NSABP trial.

As in \citet{gail1985testing}, we take three-year disease-free
survival as the outcome, so that $Y = 1$ if the subject survived
disease-free for three years after treatment, and $Y=0$ otherwise.
Patient covariates are age (years), PR (fmol), estrogen receptor level (ER, fmol), 
tumor size (centimeters), and number of histologically positive nodes
(number of nodes, integer).  We estimated the optimal treatment regime
representable as a decision list using data from the 1164 subjects
with complete observations on these variables.  Because treatment
assignment was randomized in NSABP, we estimated $\omega(x, a)$ by the sample proportion of subjects receiving treatment
$a$.  Based on exploratory analyses, we estimated $\mu(x, a)$
using a logistic regression model with transformed predictors $z = z(x) 
= \{\text{age}, \log(1 + \text{PR}), \log(1 + \text{ER}),
\text{tumor-size}, \log(1 + \text{number-of-nodes})\}^\T$.

\begin{figure}
\renewcommand{\baselinestretch}{1.1}\fontsize{11}{12}\selectfont
\centering
\vspace*{-3ex}%
\includegraphics[scale = 0.7]{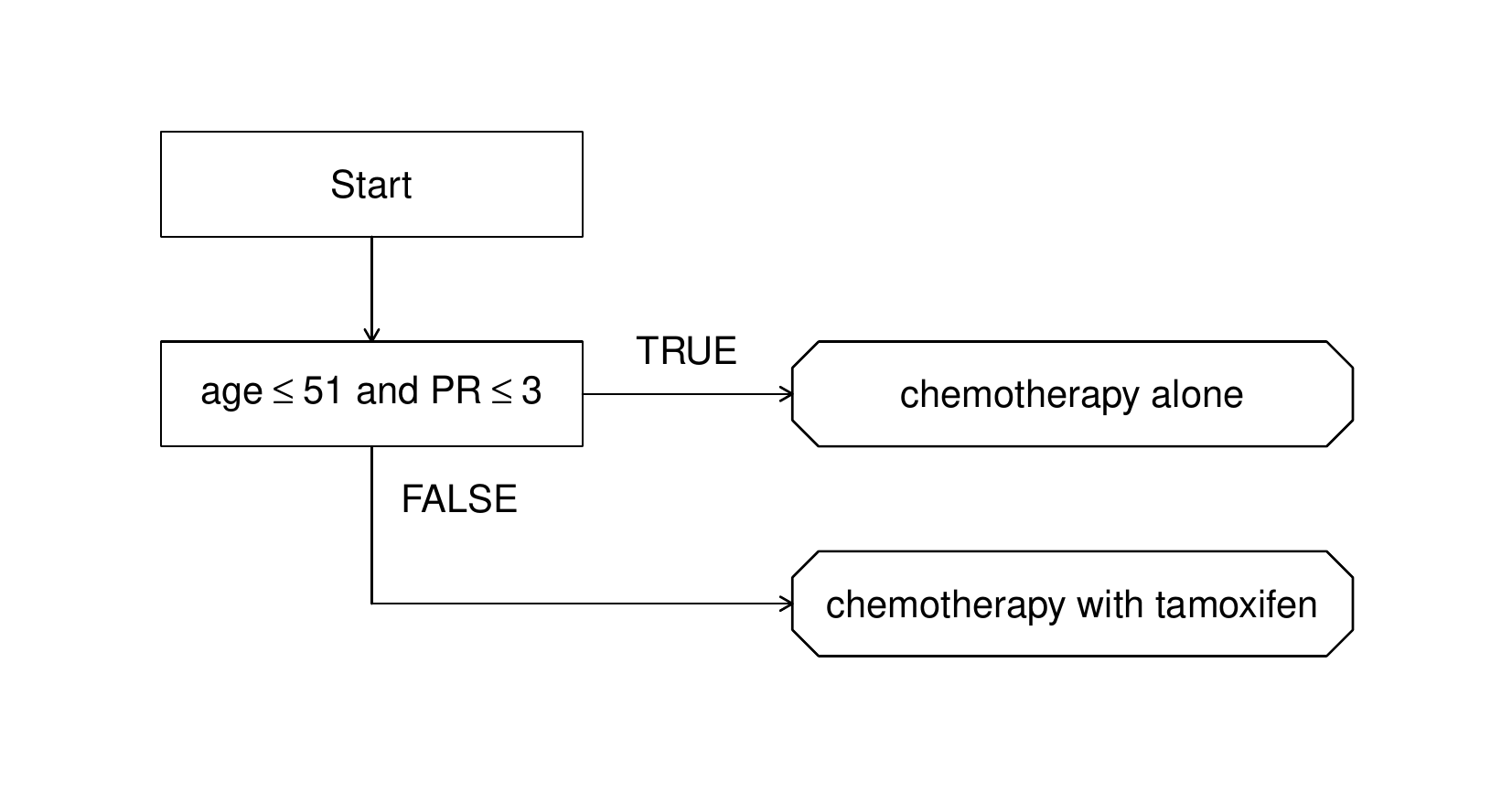}

\vspace*{-7ex}%
\includegraphics[scale = 0.7]{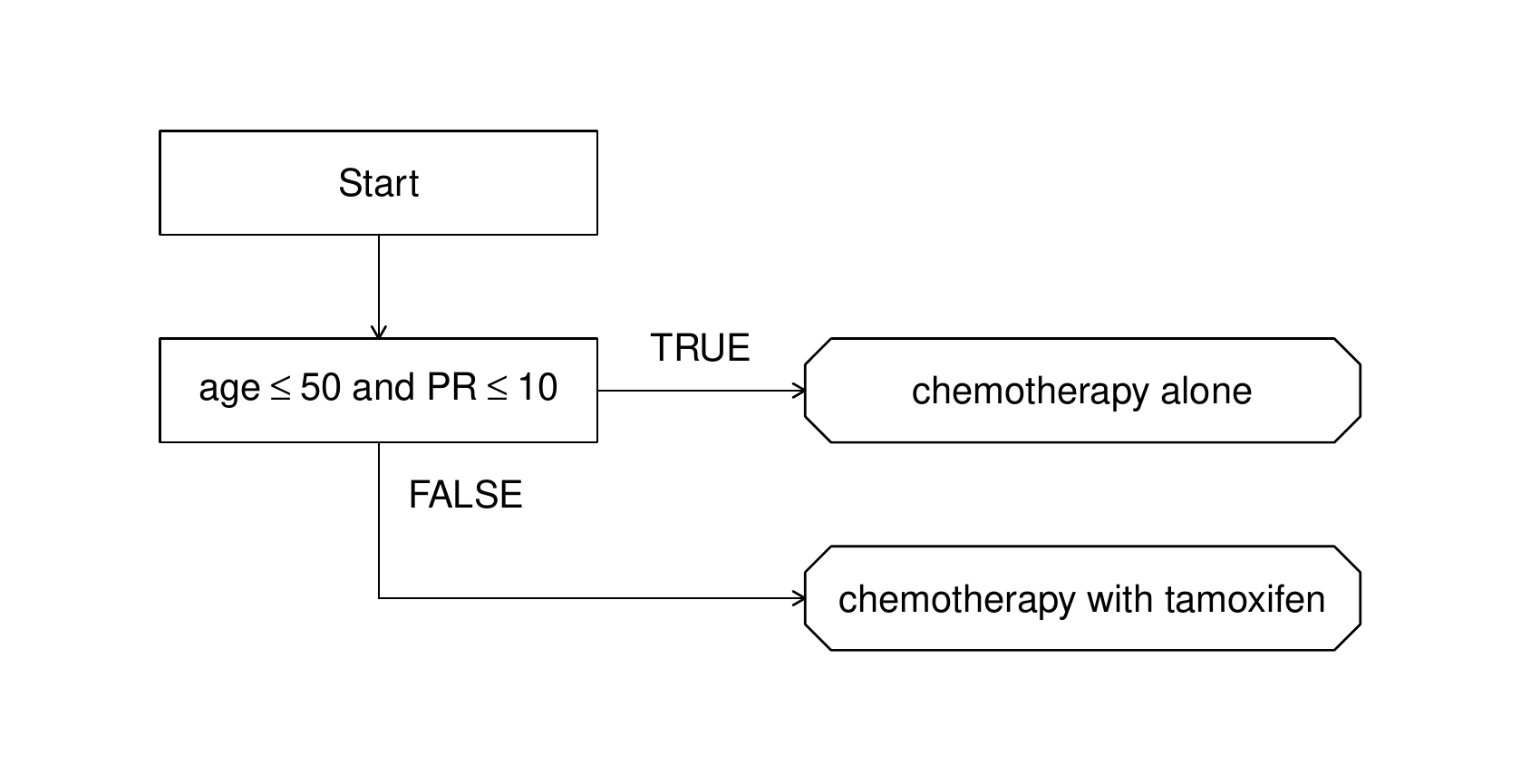}

\vspace*{-4ex}%
\caption{\textbf{Top:} estimated optimal treatment regime representable as a decision list.
\textbf{Bottom:} treatment regime proposed by \citet{gail1985testing}.}
\label{fig:app1}
\end{figure}

The estimated optimal treatment regime representable as a decision
list is given in the top panel of Figure~\ref{fig:app1}; the regime
estimated by Gail and Simon is given in the bottom panel of this
figure.  The structure of the two treatment regimes is markedly
similar.  The treatment recommendations from the two regimes agree for
92\% of the patients in the NSABP data.  In this data set, $33\%$ of
the patients have a PR value less than $3$; $13\%$ of the patients
have a PR values between $3$ and $10$; and $54\%$ of the patients have
a PR value greater than $10$.

In a previous analysis of the NSABP data, \citet{zhang2012robust}
recommended that patients with
$\text{age} + 7.98 \log(1 + \text{PR}) \leq 60$ receive chemotherapy
alone and all others receive chemotherapy plus tamoxifen.
However, this regime was built using only age and PR as potential
predictors with no data-driven variable selection.  In contrast, the
proposed method selects age and PR from the list of potential
predictors.  For completeness, we also implemented parametric
$Q$-learning using a logistic regression model with covariate vector
$z$.  The estimated regime recommends chemotherapy alone if
$ 1.674 - 0.021 \; \text{age} - 0.076 \log(1 + \text{PR}) - 0.116
\log(1 + \text{ER}) - 0.024 \;\text{tumor-size} - 0.274 \log(1 +
\text{number-of-nodes}) \geq 0 $
and chemotherapy with tamoxifen otherwise.  The treatment
recommendation dictated by parametric $Q$-learning agrees with that
dictated by decision list for $86\%$ of the subjects in the data set.

To estimate the survival probability under each estimated regime, we
use cross-validation.  The data set was
randomly divided into a training set containing $80\%$ of the subjects
and a test set containing $20\%$ of the subjects.  The optimal regime
was estimated using both approaches on the training set, and its value
was computed using~\eqref{eq:aipw2} (with $\wh{\mu} \equiv 0$) on the
test set.  To reduce variability, this process was repeated $100$
times.  The estimated survival probability is $0.65$ for the regime
representable as decision list and $0.66$ for the regime obtained from
parametric $Q$-learning.  Thus, the proposed method greatly improves
interpretability while preserving quality.

\subsection{Chronic Depression Data}

\citet{keller2000comparison} compared nefazodone, psychotherapy, and
combination of nefazodone and psychotherapy for treating
patients with chronic depression in a three-arm randomized clinical
trial. Among the three treatments considered, combination therapy was
shown to be the most beneficial in terms of efficacy as measured by
the Hamilton Rating Scale for Depression score (HRSD).  However, the 
combination treatment is significantly more expensive and burdensome 
than monotherapy. Therefore, it is of interest to construct a treatment
regime that recommends combination therapy only to subjects for whom
there is a significant benefit over monotherapy.  

Because lower HRSD indicates less severe symptoms, we define outcome $Y = 
-\text{HRSD}$ to be consistent with our paradigm of maximizing the mean 
outcome.  
Patient covariates comprise $50$ pretreatment variables, 
including personal habits and difficulties, medication history and 
various scores from several psychological questionnaires;
a list of these variables is given in the Web Appendix. 
We estimate an optimal regime using 
data from the $n=647$ (of $680$ enrolled)
subjects in the clinical trial with complete data.  
Because treatments were randomly assigned, we estimated $\omega(x, a)$ by the sample
proportion of subjects receiving treatment $a$.  We
estimated $\mu(x, a)$ using a penalized linear regression model 
with all patient covariates and treatment by
covariate interactions.  Penalization was implemented with a 
LASSO penalty tuned using 10-fold cross-validated prediction error.

The estimated optimal treatment regime representable as a decision list is displayed in
Figure~\ref{fig:example}.  One explanation for this
rule is as follows.  Those with strong physical
anxiety symptoms (somatic) and significant cognitive impairment
(retardation)
may be unlikely to benefit from psychotherapy alone or
in combination with nefazodone and are therefore recommended
to nefazodone alone.  Otherwise, because psychotherapy 
is a primary tool for treating anxiety (HAM-A) and nefazodone
is associated with sleep disturbance (sleep), it may be best to 
assign subjects with moderate to severe anxiety and severe sleep 
disturbance to psychotherapy alone.  All others
are assigned to the combination therapy. 

The estimated regime contains only four covariates.  In contrast, the
regime estimated by parametric $Q$-learning using linear regression
and LASSO penalty involves a linear combination of twenty-four
covariates, making it difficult to explain and expensive to implement.
To compare the quality of these two regimes, we use random-split
cross-validation as in Section~4.1.  The estimated HRSD score under
the regime representable as decision list is $12.9$, while that under
the regime estimated by parametric $Q$-learning is $11.8$.  Therefore,
by using decision lists we are able to obtain a remarkably more
parsimonious regime with high quality, which facilitates easier
interpretation.

\section{Discussion}
Data-driven treatment regimes have the potential to improve
patient outcomes and generate new clinical hypotheses. 
Estimation of an optimal treatment regime is typically
conducted as a secondary, exploratory analysis aimed at building
knowledge and informing future clinical research. Thus, it is
important that methodological developments are designed to fit this
exploratory role.  Decision lists are a simple yet powerful
tool for estimation of interpretable treatment regimes from
observational or experimental data.  Because decision lists
can be immediately interpreted, clinical scientists can focus on 
the scientific validity of the estimated treatment regime.  This
allows the communications between the statistician and clinical
collaborators to focus on the science rather than 
the technical details of a statistical model.

Due to the ``if-then'' format and the conditions given in \eqref{eq:cond},
the estimated regime, as a function of the data, is discrete.
Thus, a theoretical proof of the consistency of the treatment recommendations 
using decision lists is heavily technical and will be presented elsewhere.
We provide some empirical evidence in the Web Appendix 
that the estimated regime gives consistent treatment decisions.

% This work can be extended to a number of other contexts
% including multiple time points, diagnostics, and risk prediction.
% We believe that these extension are important to facilitate
% better information transfer and shared decision between
% patients and their healthcare providers.  Furthermore,
% short-circuited treatment recommendations or outcome
% predictions have the potential to reduce patient 
% cost and burden.

% The proposed approach can be extended to accommodate 
% multiple decision points.
% We are currently developing an approach, 
% and will report on it in a future article.
% Another potential advantage of decision lists is that they allow for
% short-circuited treatment decisions which can be used to reduce
% patient burden and costs.  We believe that this is an important consideration in
% contexts where patient measurements are costly, time consuming, or
% burdensome.  It would be interesting to extend these
% ideas to other contexts, including risk prediction and
% disease diagnostics.

\section*{Acknowledgments}
This work was supported by NIH grants P01 CA142538, R01 CA085848, and R01 HL118336.
The authors thank the NSABP for providing the clinical trial data,
and thank John Rush for providing the chronic depression data.

\section*{Supplementary Materials}

Web Appendix, referenced in Sections~2.4.1, 2.4.2, 3 and 4.2, 
is available with this paper.

\bibliographystyle{apalike}
\bibliography{references}

\clearpage

\appendix

\begin{center}
\Large
Web Appendix for ``Using decision lists to construct interpretable and parsimonious
treatment regimes''
\end{center}

\FloatBarrier

\setcounter{section}{0}
\renewcommand{\thesection}{A.\arabic{section}}

\section{An Illustrative Run Through the Algorithm for Finding an Optimal Decision List}

In this section, we illustrate how the proposed algorithm for finding an optimal decision list works. 
For simplicity, the patient covariate is assumed to be two-dimensional.

\begin{itemize}
\item The algorithm starts at Step~1.

\begin{itemize}
\item We choose $L_{\max} = 5$ and $\alpha = 0.05$.

\item We compute $\wt{a}_{0} = \argmax_{a_0\in\mathcal{A}} \wh{R}\left[ \lb a_0 \rb \right] $. 
Suppose the maximum found is $\wh{R}\left[ \lb \wt{a}_0 \rb \right] = 10$.
Figure~\ref{fig:step1.2} shows the decision list $ \lb \wt{a}_0 \rb $.

\item We set $\Pi_\text{temp} = \emptyset$ and $\Pi_\text{final} = \emptyset$. 
\end{itemize}

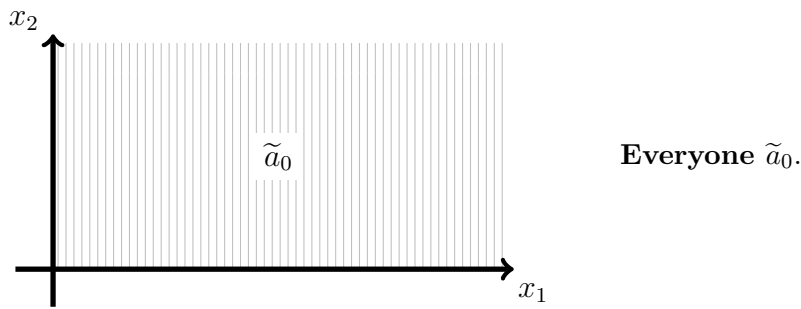
\begin{figure}
\renewcommand{\baselinestretch}{1.1}\fontsize{11}{12}\selectfont

\hfill
\begin{minipage}{0.4\linewidth}
\centering
\newcommand{\lnth}{1pt} %\scalefont{50} %
\begin{tikzpicture}[scale=0.5, every node/.style={font=\normalsize}]

\pattern[pattern=vertical lines, pattern color=lightgray] (0, 0) rectangle (12, 6);
\node at (6, 3) [fill=white] {$\wt{a}_0$};

%% Thick lines
\draw[line width=2*\lnth, ->] (0, -1) -- (0, 6.25);
\draw[line width=2*\lnth, ->] (-1, 0) -- (12.25, 0);

%% Axis labels
\node at (-0.1, 6.1) [anchor=south east] {$x_2$};
\node at (12.1, -0.1) [anchor=north west] {$x_1$};

\end{tikzpicture}
\end{minipage}%
\hspace{0.1\linewidth}%
\begin{minipage}{0.4\linewidth}
$\textbf{Everyone } \wt{a}_0$. 
\end{minipage}%
\hfill

\caption{ 
Diagram and description of the decision list $ \lb \wt{a}_0 \rb $. 
}
\label{fig:step1.2}
\end{figure}

\item The algorithm proceeds to Step~2. 

\begin{itemize}
\item The goal is to estimate the first clause $(c_1, a_1)$. 

\item We compute $ 
  (\wt{c}_{1}, \wt{a}_{1}, \wt{a}_{1}') = 
  \argmax_{(c_{1}, a_{1}, a_{1}') \in \mathcal{C} \times \mathcal{A} \times \mathcal{A}}
  \wh{R}\left[ \lb (c_{1}, a_{1}), a_{1}' \rb \right]
$. This is done, conceptually, by evaluating $\wh{R}(\cdot)$ at 
each element in $\mathcal{C} \times \mathcal{A} \times \mathcal{A}$. 
Suppose the maximum found is $\wh{R}\left[ \lb (\wt{c}_1, \wt{a}_1), \wt{a}_1' \rb \right] = 15$
and the clause $\wt{c}_1$ has the form $x_1 \leq \tau_1$.
Figure~\ref{fig:step2.1} shows the decision list $ \lb (\wt{c}_{1}, \wt{a}_{1}), \wt{a}_{1}' \rb $.

\item We compute $ 
  \wh{\Delta}_1 = \wh{R}\left[ \lb (\wt{c}_{1}, \wt{a}_{1}), \wt{a}_{1}' \rb \right] - \wh{R}\left[ \lb \wt{a}_0 \rb \right]
$ and compare $\wh{\Delta}_1$ to 
  $z_{1-\alpha} \big\{\wh{\mathrm{Var}} \big(\wh{\Delta}_1\big)\big\}^{1/2}$.
In this case $\wh{\Delta}_1 = 15 - 10 = 5$.
Suppose we get $\wh{\mathrm{Var}} \big(\wh{\Delta}_1\big) = 4$ after calculations.
Since $5 > z_{0.95} \times 4^{1/2}$, 
we add two decision lists, $ \lb (\wt{c}_{1}, \wt{a}_{1}), \wt{a}_{1}' \rb $ and 
$ \lb (\wt{c}_{1}', \wt{a}_{1}'), \wt{a}_{1} \rb $,
into the set $\Pi_\text{temp}$ and proceed to estimate the second clause $(c_2, a_2)$.

\item We make a remark on non-uniqueness here.
The decision list $ \lb (\wt{c}_{1}, \wt{a}_{1}), \wt{a}_{1}' \rb $ can be equivalently expressed as $ \lb (\wt{c}_{1}', \wt{a}_{1}'), \wt{a}_{1} \rb $,
where $\wt{c}_{1}'$ is the negation of $\wt{c}_{1}$.
Since these two decision lists provide the same treatment recommendation to every patient,
we have $\wh{R}[ \lb (\wt{c}_{1}', \wt{a}_{1}'), \wt{a}_{1} \rb ]
= \wh{R} [ \lb (\wt{c}_{1}, \wt{a}_{1}), \wt{a}_{1}' \rb ] = 15$.
However, their first clauses are different and may lead to considerably different final decision lists. 
Currently it is impossible to determine whether $(\wt{c}_1, \wt{a}_1)$ or $(\wt{c}_1', \wt{a}_1')$ 
should be used in the first clause.
Thus we add both decision lists into $\Pi_\text{temp}$, and
move on to building the second clause while keeping in mind that there are two possibilities, $(\wt{c}_1, \wt{a}_1)$ and $(\wt{c}_1', \wt{a}_1')$, for the first clause. 
Figure~\ref{fig:step2.r} shows the decision list $ \lb (\wt{c}_{1}', \wt{a}_{1}'), \wt{a}_{1} \rb $.
The diagram is the same as in Figure~\ref{fig:step2.1} while the description is different.
\end{itemize}

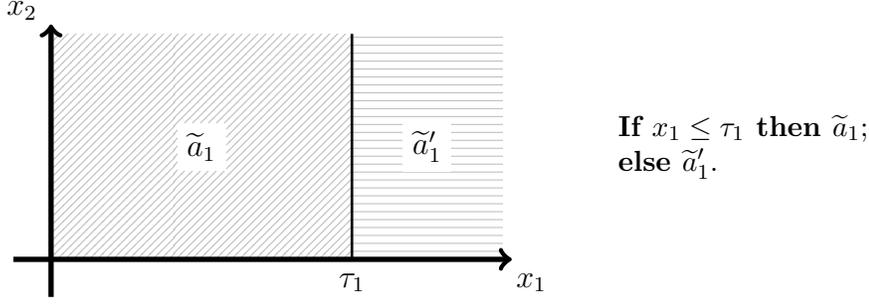
\begin{figure}
\renewcommand{\baselinestretch}{1.1}\fontsize{11}{12}\selectfont

\hfill
\begin{minipage}{0.4\linewidth}
\centering
\newcommand{\lnth}{1pt} %\scalefont{50} %
\begin{tikzpicture}[scale=0.5, every node/.style={font=\normalsize}]

\pattern[pattern=north east lines, pattern color=lightgray] (0, 0) rectangle (8, 6);
\node at (4, 3) [fill=white] {$\wt{a}_1$};

\pattern[pattern=horizontal lines, pattern color=lightgray] (8, 0) rectangle (12, 6);
\node at (10, 3) [fill=white] {$\wt{a}_1'$};

%% Thick lines
\draw[line width=\lnth] (8, 0) -- (8, 6);
\draw[line width=2*\lnth, ->] (0, -1) -- (0, 6.25);
\draw[line width=2*\lnth, ->] (-1, 0) -- (12.25, 0);

%% Axis labels
\node at (8, -0.1) [anchor=north] {$\tau_1$};
\node at (-0.1, 6.1) [anchor=south east] {$x_2$};
\node at (12.1, -0.1) [anchor=north west] {$x_1$};

\end{tikzpicture}
\end{minipage}%
\hspace{0.1\linewidth}%
\begin{minipage}{0.4\linewidth}
$\textbf{If } x_1 \leq \tau_1 \textbf{ then } \wt{a}_1$; \\ 
$\textbf{else } \wt{a}_1'$. 
\end{minipage}%
\hfill

\caption{ 
Diagram and description of the decision list $ \lb (\wt{c}_{1}, \wt{a}_{1}), \wt{a}_{1}' \rb $. 
}
\label{fig:step2.1}
\end{figure}

\begin{figure}
\renewcommand{\baselinestretch}{1.1}\fontsize{11}{12}\selectfont

\hfill
\begin{minipage}{0.4\linewidth}
\centering
\newcommand{\lnth}{1pt} %\scalefont{50} %
\begin{tikzpicture}[scale=0.5, every node/.style={font=\normalsize}]

\pattern[pattern=north east lines, pattern color=lightgray] (0, 0) rectangle (8, 6);
\node at (4, 3) [fill=white] {$\wt{a}_1$};

\pattern[pattern=horizontal lines, pattern color=lightgray] (8, 0) rectangle (12, 6);
\node at (10, 3) [fill=white] {$\wt{a}_1'$};

%% Thick lines
\draw[line width=\lnth] (8, 0) -- (8, 6);
\draw[line width=2*\lnth, ->] (0, -1) -- (0, 6.25);
\draw[line width=2*\lnth, ->] (-1, 0) -- (12.25, 0);

%% Axis labels
\node at (8, -0.1) [anchor=north] {$\tau_1$};
\node at (-0.1, 6.1) [anchor=south east] {$x_2$};
\node at (12.1, -0.1) [anchor=north west] {$x_1$};

\end{tikzpicture}
\end{minipage}%
\hspace{0.1\linewidth}%
\begin{minipage}{0.4\linewidth}
$\textbf{If } x_1 > \tau_1 \textbf{ then } \wt{a}_1'$; \\ 
$\textbf{else } \wt{a}_1$. 
\end{minipage}%
\hfill

\caption{ 
Diagram and description of the decision list $ \lb (\wt{c}_{1}', \wt{a}_{1}'), \wt{a}_{1} \rb $. 
}
\label{fig:step2.r}
\end{figure}

\item The algorithm proceeds to Step~3.

\begin{itemize}
\item We pick an element $\wb{\pi}$ from $\Pi_\text{temp}$.
Currently $\Pi_\text{temp}$ contains two decision lists:
$ \lb (\wt{c}_{1}, \wt{a}_{1}), \wt{a}_{1}' \rb $ and 
$ \lb (\wt{c}_{1}', \wt{a}_{1}'), \wt{a}_{1} \rb $.
Suppose we get 
$\wb{\pi} = \lb (\wt{c}_{1}, \wt{a}_{1}), \wt{a}_{1}' \rb $.
We remove $\wb{\pi}$ from $\Pi_\text{temp}$.

\item We compute $
(\wt{c}_{2}, \wt{a}_{2}, \wt{a}_{2}') = 
    \argmax_{(c_{2}, a_{2}, a_{2}') \in \mathcal{C} \times \mathcal{A} \times \mathcal{A}} 
    \wh{R}\left[ \lb (\wt{c}_1, \wt{a}_1), (c_{2}, a_{2}), a_{2}' \rb \right]
$. During the maximization $(\wt{c}_1, \wt{a}_1)$ is held fixed.
Intuitively, this is to partition $\mathcal{T}(\wt{c}_1)^c$
while keeping $\mathcal{T}(\wt{c}_1)$ fixed.
Suppose the maximum found is $\wh{R} [ \lb (\wt{c}_1, \wt{a}_1), (\wt{c}_{2}, \wt{a}_{2}), \wt{a}_{2}' \rb ] = 16$ 
and the clause $\wt{c}_2$ has the form $x_2 \leq \tau_{21}$.
Figure~\ref{fig:step3.1} shows the decision list $ \lb (\wt{c}_1, \wt{a}_1), (\wt{c}_{2}, \wt{a}_{2}), \wt{a}_{2}' \rb $.

\item We compute $ 
\wh{\Delta}_2 = 
    \wh{R}\left[ \lb (\wt{c}_1, \wt{a}_1), (\wt{c}_{2}, \wt{a}_{2}), \wt{a}_{2}' \rb \right] 
    - \wh{R}\left[ \lb (\wt{c}_1, \wt{a}_1), \wt{a}_1' \rb \right] 
$ and compare $\wh{\Delta}_2$ to\break
  $z_{1-\alpha} \big\{\wh{\mathrm{Var}} \big(\wh{\Delta}_2\big)\big\}^{1/2}$.
In this case $\wh{\Delta}_2 = 16 - 15 = 1$.
Suppose we get $\wh{\mathrm{Var}} \big(\wh{\Delta}_2\big) = 2.25$ after calculations.
Since $\wh{\Delta}_2 < z_{0.95} \big\{\wh{\mathrm{Var}} \big(\wh{\Delta}_2\big)\big\}^{1/2}$,
the simpler, more parsimonious decision list $\lb (\wt{c}_1, \wt{a}_1), \wt{a}_1' \rb$ is preferred
and added to $\Pi_\text{final}$, while $ \lb (\wt{c}_1, \wt{a}_1), (\wt{c}_{2}, \wt{a}_{2}), \wt{a}_{2}' \rb $ is discarded.
\end{itemize}

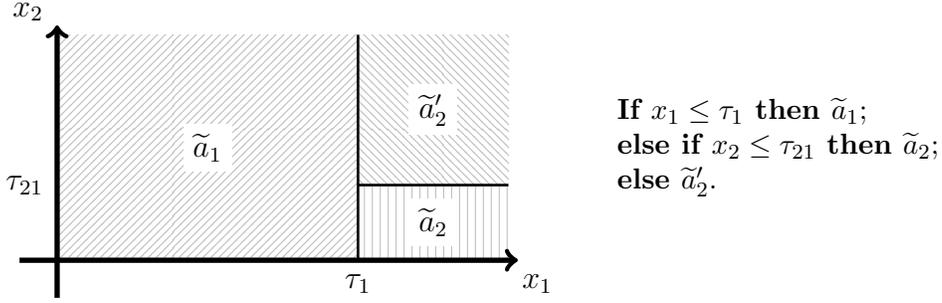
\begin{figure}
\renewcommand{\baselinestretch}{1.1}\fontsize{11}{12}\selectfont

\hfill
\begin{minipage}{0.4\linewidth}
\centering
\newcommand{\lnth}{1pt} %\scalefont{50} %
\begin{tikzpicture}[scale=0.5, every node/.style={font=\normalsize}]

\pattern[pattern=north east lines, pattern color=lightgray] (0, 0) rectangle (8, 6);
\node at (4, 3) [fill=white] {$\wt{a}_1$};

\pattern[pattern=vertical lines, pattern color=lightgray] (8, 0) rectangle (12, 2);
\node at (10, 1) [fill=white] {$\wt{a}_2$};

\pattern[pattern=north west lines, pattern color=lightgray] (8, 2) rectangle (12, 6);
\node at (10, 4) [fill=white] {$\wt{a}_2'$};

%% Thick lines
\draw[line width=\lnth] (8, 2) -- (12, 2);
\draw[line width=\lnth] (8, 0) -- (8, 6);
\draw[line width=2*\lnth, ->] (0, -1) -- (0, 6.25);
\draw[line width=2*\lnth, ->] (-1, 0) -- (12.25, 0);

%% Axis labels
\node at (-0.1, 2) [anchor=east] {$\tau_{21}$};
\node at (8, -0.1) [anchor=north] {$\tau_1$};
\node at (-0.1, 6.1) [anchor=south east] {$x_2$};
\node at (12.1, -0.1) [anchor=north west] {$x_1$};

\end{tikzpicture}
\end{minipage}%
\hspace{0.1\linewidth}%
\begin{minipage}{0.4\linewidth}
$\textbf{If } x_1 \leq \tau_1 \textbf{ then } \wt{a}_1$; \\ 
$\textbf{else if } x_2 \leq \tau_{21} \textbf{ then } \wt{a}_2$; \\ 
$\textbf{else } \wt{a}_2'$. 
\end{minipage}%
\hfill

\caption{ 
Diagram and description of the decision list $ \lb (\wt{c}_1, \wt{a}_1), (\wt{c}_{2}, \wt{a}_{2}), \wt{a}_{2}' \rb $. 
It is possible that $\wt{a}_2 = \wt{a}_1$ or $\wt{a}_2' = \wt{a}_1$.
}
\label{fig:step3.1}
\end{figure}

\item The algorithm repeats Step~3.

\begin{itemize}
\item Step~3 is repeated since $\Pi_\text{temp}$ contains another element 
$\wb{\pi} = \lb (\wt{c}_1', \wt{a}_1'), \wt{a}_1 \rb$.
We remove $\wb{\pi}$ from $\Pi_\text{temp}$.

\item We compute $
(\wt{c}_{2}, \wt{a}_{2}, \wt{a}_{2}') = 
    \argmax_{(c_{2}, a_{2}, a_{2}') \in \mathcal{C} \times \mathcal{A} \times \mathcal{A}} 
    \wh{R}\left[ \lb (\wt{c}_1', \wt{a}_1'), (c_{2}, a_{2}), a_{2}' \rb \right]
$. During the maximization $(\wt{c}_1', \wt{a}_1')$ is held fixed.
Intuitively, this is to partition $\mathcal{T}(\wt{c}_1)$
while keeping $\mathcal{T}(\wt{c}_1)^c$ fixed.
Suppose the maximum found is $\wh{R} [\lb (\wt{c}_1', \wt{a}_1'), (\wt{c}_{2}, \wt{a}_{2}), \wt{a}_{2}' \rb)] = 18$ 
and the clause $\wt{c}_2$ has the form $x_2 \leq \tau_{22}$.
Figure~\ref{fig:step4.1} shows the decision list $ \lb (\wt{c}_1', \wt{a}_1'), (\wt{c}_{2}, \wt{a}_{2}), \wt{a}_{2}' \rb $.

\item We compute $ 
\wh{\Delta}_2 = 
    \wh{R}\left[ \lb (\wt{c}_1', \wt{a}_1'), (\wt{c}_{2}, \wt{a}_{2}), \wt{a}_{2}' \rb \right] 
    - \wh{R}\left[ \lb (\wt{c}_1', \wt{a}_1'), \wt{a}_1 \rb \right] 
$ and compare $\wh{\Delta}_2$ to\break
  $z_{1-\alpha} \big\{\wh{\mathrm{Var}} \big(\wh{\Delta}_2\big)\big\}^{1/2}$.
In this case $\wh{\Delta}_2 = 18 - 15 = 3$.
Suppose we get $\wh{\mathrm{Var}} \big(\wh{\Delta}_2\big) = 2$ after calculations.
Then we have $\wh{\Delta}_2 > z_{0.95} \big\{\wh{\mathrm{Var}} \big(\wh{\Delta}_2\big)\big\}^{1/2}$,
which means that the second clause significantly improves the performance of the decision list.
Thus we add decision lists $ \lb (\wt{c}_1', \wt{a}_1'), (\wt{c}_{2}, \wt{a}_{2}), \wt{a}_{2}' \rb $
and $ \lb (\wt{c}_1', \wt{a}_1'), (\wt{c}_{2}', \wt{a}_{2}'), \wt{a}_{2} \rb $ 
to $\Pi_\text{temp}$.

\item Here the non-uniqueness comes into play again. Consequently,
although the decision lists $ \lb (\wt{c}_1', \wt{a}_1'), (\wt{c}_{2}, \wt{a}_{2}), \wt{a}_{2}' \rb $
and $ \lb (\wt{c}_1', \wt{a}_1'), (\wt{c}_{2}', \wt{a}_{2}'), \wt{a}_{2} \rb $
are equivalent,
it is important to have both of them added to $\Pi_\text{temp}$.
\end{itemize}

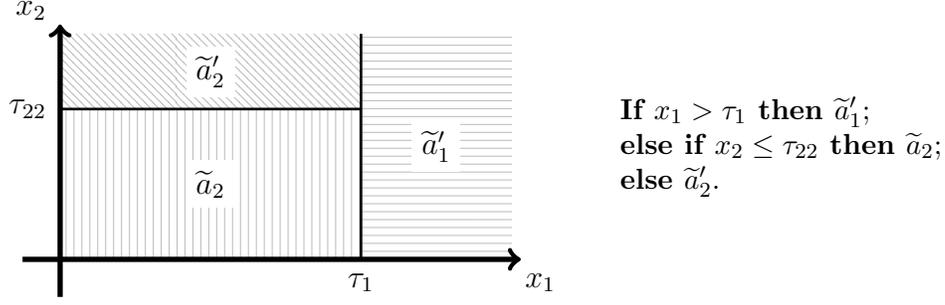
\begin{figure}
\renewcommand{\baselinestretch}{1.1}\fontsize{11}{12}\selectfont

\hfill
\begin{minipage}{0.4\linewidth}
\centering
\newcommand{\lnth}{1pt} %\scalefont{50} %
\begin{tikzpicture}[scale=0.5, every node/.style={font=\normalsize}]

\pattern[pattern=horizontal lines, pattern color=lightgray] (8, 0) rectangle (12, 6);
\node at (10, 3) [fill=white] {$\wt{a}_1'$};

\pattern[pattern=vertical lines, pattern color=lightgray] (0, 0) rectangle (8, 4);
\node at (4, 2) [fill=white] {$\wt{a}_2$};

\pattern[pattern=north west lines, pattern color=lightgray] (0, 4) rectangle (8, 6);
\node at (4, 5) [fill=white] {$\wt{a}_2'$};

%% Thick lines
\draw[line width=\lnth] (0, 4) -- (8, 4);
\draw[line width=\lnth] (8, 0) -- (8, 6);
\draw[line width=2*\lnth, ->] (0, -1) -- (0, 6.25);
\draw[line width=2*\lnth, ->] (-1, 0) -- (12.25, 0);

%% Axis labels
\node at (-0.1, 4) [anchor=east] {$\tau_{22}$};
\node at (8, -0.1) [anchor=north] {$\tau_1$};
\node at (-0.1, 6.1) [anchor=south east] {$x_2$};
\node at (12.1, -0.1) [anchor=north west] {$x_1$};

\end{tikzpicture}
\end{minipage}%
\hspace{0.1\linewidth}%
\begin{minipage}{0.4\linewidth}
$\textbf{If } x_1 > \tau_1 \textbf{ then } \wt{a}_1'$; \\ 
$\textbf{else if } x_2 \leq \tau_{22} \textbf{ then } \wt{a}_2$; \\ 
$\textbf{else } \wt{a}_2'$. 
\end{minipage}%
\hfill

\caption{ 
Diagram and description of the decision list $ \lb (\wt{c}_1', \wt{a}_1'), (\wt{c}_{2}, \wt{a}_{2}), \wt{a}_{2}' \rb $. 
It is possible that $\wt{a}_2 = \wt{a}_1'$ or $\wt{a}_2' = \wt{a}_1'$.
}
\label{fig:step4.1}
\end{figure}

\item The algorithm repeats Step~3.

\begin{itemize}
\item Now $\Pi_\text{temp}$ contains two decision lists while $\Pi_\text{final}$ contains one. 
Thus Step~3 is repeated.
We first pick and remove an element $\wb{\pi}$ from $\Pi_\text{temp}$,
say $ \wb{\pi} = \lb (\wt{c}_1', \wt{a}_1'), (\wt{c}_{2}, \wt{a}_{2}), \wt{a}_{2}' \rb $.

\item Next, we will build a decision list of length $3$ and the first two clauses being 
$ (\wt{c}_1', \wt{a}_1')$ and $(\wt{c}_{2}, \wt{a}_{2})$.
We compute $
(\wt{c}_{3}, \wt{a}_{3}, \wt{a}_{3}') = 
    \argmax_{(c_{3}, a_{3}, a_{3}') \in \mathcal{C} \times \mathcal{A} \times \mathcal{A}} 
    \wh{R}\left[ \lb (\wt{c}_1', \wt{a}_1'), (\wt{c}_{2}, \wt{a}_{2}), (c_{3}, a_{3}), a_{3}' \rb \right]
$. During the maximization $(\wt{c}_1', \wt{a}_1')$ and $(\wt{c}_2, \wt{a}_2)$ are held fixed.
Suppose the maximum found is $\wh{R} [ \lb (\wt{c}_1', \wt{a}_1'), (\wt{c}_{2}, \wt{a}_{2}), (\wt{c}_{3}, \wt{a}_{3}), \wt{a}_{3}' \rb] = 20$ and the clause $\wt{c}_3$ has the form $x_1 \leq \tau_{31}$.
Figure~\ref{fig:step5.1} shows the decision list $ \lb (\wt{c}_1', \wt{a}_1'), (\wt{c}_{2}, \wt{a}_{2}), (\wt{c}_{3}, \wt{a}_{3}), \wt{a}_{3}' \rb $.

\item We then compute $ 
\wh{\Delta}_3 = 
    \wh{R}\left[ \lb (\wt{c}_1', \wt{a}_1'), (\wt{c}_{2}, \wt{a}_{2}), (\wt{c}_{3}, \wt{a}_{3}), \wt{a}_{3}' \rb \right]
    - \wh{R}\left[ \lb (\wt{c}_1', \wt{a}_1'), (\wt{c}_{2}, \wt{a}_{2}), \wt{a}_{2}' \rb \right] 
$ and compare $\wh{\Delta}_3$ to 
  $z_{1-\alpha} \big\{\wh{\mathrm{Var}} \big(\wh{\Delta}_3\big)\big\}^{1/2}$.
In this case $\wh{\Delta}_3 = 20 - 18 = 2$.
Suppose we get $\wh{\mathrm{Var}} \big(\wh{\Delta}_3\big) = 3$ after calculations.
Then we have $\wh{\Delta}_3 < z_{0.95} \big\{\wh{\mathrm{Var}} \big(\wh{\Delta}_3\big)\big\}^{1/2}$.
Thus the simpler, more parsimonious, decision list $ \lb (\wt{c}_1', \wt{a}_1'), (\wt{c}_{2}, \wt{a}_{2}), \wt{a}_{2}' \rb $ is preferred.
So we add $ \lb (\wt{c}_1', \wt{a}_1'), (\wt{c}_{2}, \wt{a}_{2}), \wt{a}_{2}' \rb $ to $\Pi_\text{final}$
and drop $ \lb (\wt{c}_1', \wt{a}_1'), (\wt{c}_{2}, \wt{a}_{2}), (\wt{c}_{3}, \wt{a}_{3}), \wt{a}_{3}' \rb $.
\end{itemize}

\begin{figure}
\renewcommand{\baselinestretch}{1.1}\fontsize{11}{12}\selectfont

\hfill
\begin{minipage}{0.4\linewidth}
\centering
\newcommand{\lnth}{1pt} %\scalefont{50} %
\begin{tikzpicture}[scale=0.5, every node/.style={font=\normalsize}]

\pattern[pattern=horizontal lines, pattern color=lightgray] (8, 0) rectangle (12, 6);
\node at (10, 3) [fill=white] {$\wt{a}_1'$};

\pattern[pattern=vertical lines, pattern color=lightgray] (0, 0) rectangle (8, 4);
\node at (4, 2) [fill=white] {$\wt{a}_2$};

\pattern[pattern=bricks, pattern color=lightgray] (0, 4) rectangle (4, 6);
\node at (2, 5) [fill=white] {$\wt{a}_3$};

\pattern[pattern=north east lines, pattern color=lightgray] (4, 4) rectangle (8, 6);
\node at (6, 5) [fill=white] {$\wt{a}_3'$};

%% Thick lines
\draw[line width=\lnth] (0, 4) -- (8, 4);
\draw[line width=\lnth] (8, 0) -- (8, 6);
\draw[line width=\lnth] (4, 4) -- (4, 6);
\draw[line width=2*\lnth, ->] (0, -1) -- (0, 6.25);
\draw[line width=2*\lnth, ->] (-1, 0) -- (12.25, 0);

%% Axis labels
\node at (-0.1, 4) [anchor=east] {$\tau_{22}$};
\node at (8, -0.1) [anchor=north] {$\tau_1$};
\node at (4, -0.1) [anchor=north] {$\tau_{31}$};
\node at (-0.1, 6.1) [anchor=south east] {$x_2$};
\node at (12.1, -0.1) [anchor=north west] {$x_1$};

\end{tikzpicture}
\end{minipage}%
\hspace{0.1\linewidth}%
\begin{minipage}{0.4\linewidth}
$\textbf{If } x_1 > \tau_1 \textbf{ then } \wt{a}_1'$; \\ 
$\textbf{else if } x_2 \leq \tau_{22} \textbf{ then } \wt{a}_2$; \\ 
$\textbf{else if } x_1 \leq \tau_{31} \textbf{ then } \wt{a}_3$; \\ 
$\textbf{else } \wt{a}_3'$. 
\end{minipage}%
\hfill

\caption{ 
Diagram and description of the decision list $ \lb (\wt{c}_1', \wt{a}_1'), (\wt{c}_{2}, \wt{a}_{2}), (\wt{c}_{3}, \wt{a}_{3}), \wt{a}_{3}' \rb $. 
Some of the values of $\wt{a}_1', \wt{a}_2, \wt{a}_3, \wt{a}_3'$ can be equal.
}
\label{fig:step5.1}
\end{figure}
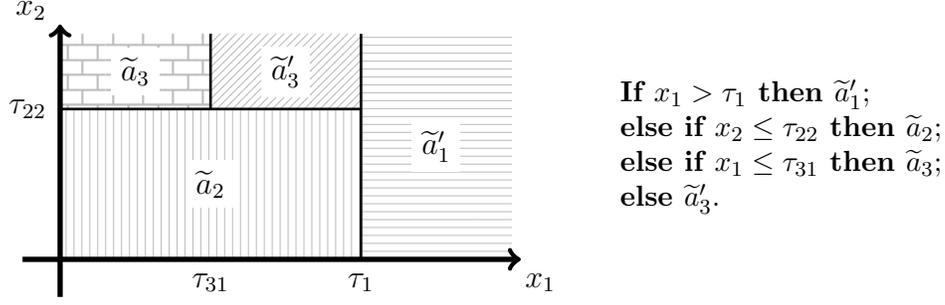

\item The algorithm repeats Step~3.
\begin{itemize}
\item Since $\Pi_\text{temp}$ contains one element 
$ \wb{\pi} = \lb (\wt{c}_1', \wt{a}_1'), (\wt{c}_{2}', \wt{a}_{2}'), \wt{a}_{2} \rb $,
we repeat Step~3 once again.
We remove $ \wb{\pi} $ from $\Pi_\text{temp}$.

\item We compute $
(\wt{c}_{3}, \wt{a}_{3}, \wt{a}_{3}') = 
    \argmax_{(c_{3}, a_{3}, a_{3}') \in \mathcal{C} \times \mathcal{A} \times \mathcal{A}} 
    \wh{R}\left[ \lb (\wt{c}_1', \wt{a}_1'), (\wt{c}_{2}', \wt{a}_{2}'), (c_{3}, a_{3}), a_{3}' \rb \right]
$ while keeping $(\wt{c}_1', \wt{a}_1')$ and $(\wt{c}_2', \wt{a}_2')$ fixed.
Suppose 
$\wh{R} ( \lb (\wt{c}_1', \wt{a}_1'), (\wt{c}_{2}', \wt{a}_{2}'), (\wt{c}_{3}, \wt{a}_{3}), \wt{a}_{3}' \rb) = 20.5$ and the clause $\wt{c}_3$ has the form $x_1 \leq \tau_{32} \text{ and } x_2 > \tau_{33}$.
Figure~\ref{fig:step6.1} shows the decision list $ \lb (\wt{c}_1', \wt{a}_1'), (\wt{c}_{2}', \wt{a}_{2}'), (\wt{c}_{3}, \wt{a}_{3}), \wt{a}_{3}' \rb $.

\item We compute $ 
\wh{\Delta}_3 = 
    \wh{R}\left[ \lb (\wt{c}_1', \wt{a}_1'), (\wt{c}_{2}', \wt{a}_{2}'), (\wt{c}_{3}, \wt{a}_{3}), \wt{a}_{3}' \rb \right] 
    - \wh{R}\left[ \lb (\wt{c}_1', \wt{a}_1'), (\wt{c}_{2}', \wt{a}_{2}'), \wt{a}_{2} \rb \right] 
$ and compare $\wh{\Delta}_3$ to 
  $z_{1-\alpha} \big\{\wh{\mathrm{Var}} \big(\wh{\Delta}_3\big)\big\}^{1/2}$.
In this case $\wh{\Delta}_1 = 20.5 - 18 = 2.5$.
Suppose we get $\wh{\mathrm{Var}} \big(\wh{\Delta}_1\big) = 2.56$ after calculations.
Then we have $\wh{\Delta}_3 < z_{0.95} \big\{\wh{\mathrm{Var}} \big(\wh{\Delta}_3\big)\big\}^{1/2}$.
So the simpler, more parsimonious, decision list $ \lb (\wt{c}_1', \wt{a}_1'), (\wt{c}_{2}', \wt{a}_{2}'), \wt{a}_{2} \rb $ is preferred.
Consequently, we add $ \lb (\wt{c}_1', \wt{a}_1'), (\wt{c}_{2}', \wt{a}_{2}'), \wt{a}_{2} \rb $ to $\Pi_\text{final}$
and discard $ \lb (\wt{c}_1', \wt{a}_1'), (\wt{c}_{2}', \wt{a}_{2}'), (\wt{c}_{3}, \wt{a}_{3}), \wt{a}_{3}' \rb $.
\end{itemize}

\begin{figure}
\renewcommand{\baselinestretch}{1.1}\fontsize{11}{12}\selectfont

\hfill
\begin{minipage}{0.4\linewidth}
\centering
\newcommand{\lnth}{1pt} %\scalefont{50} %
\begin{tikzpicture}[scale=0.5, every node/.style={font=\normalsize}]

\pattern[pattern=horizontal lines, pattern color=lightgray] (8, 0) rectangle (12, 6);
\node at (10, 3) [fill=white] {$\wt{a}_1'$};

\pattern[pattern=north west lines, pattern color=lightgray] (0, 4) rectangle (8, 6);
\node at (4, 5) [fill=white] {$\wt{a}_2'$};

\pattern[pattern=grid, pattern color=lightgray] (0, 2) rectangle (4, 4);
\node at (2, 3) [fill=white] {$\wt{a}_3$};

\pattern[pattern=bricks, pattern color=lightgray] (0, 0) rectangle (8, 2) rectangle (4, 4);
\node at (6, 2) [fill=white] {$\wt{a}_3'$};

%% Thick lines
\draw[line width=\lnth] (0, 4) -- (8, 4);
\draw[line width=\lnth] (8, 0) -- (8, 6);
\draw[line width=\lnth] (0, 2) -- (4, 2) -- (4, 4);
\draw[line width=2*\lnth, ->] (0, -1) -- (0, 6.25);
\draw[line width=2*\lnth, ->] (-1, 0) -- (12.25, 0);

%% Axis labels
\node at (-0.1, 4) [anchor=east] {$\tau_{22}$};
\node at (8, -0.1) [anchor=north] {$\tau_1$};
\node at (-0.1, 2) [anchor=east] {$\tau_{33}$};
\node at (4, -0.1) [anchor=north] {$\tau_{32}$};
\node at (-0.1, 6.1) [anchor=south east] {$x_2$};
\node at (12.1, -0.1) [anchor=north west] {$x_1$};

\end{tikzpicture}
\end{minipage}%
\hspace{0.1\linewidth}%
\begin{minipage}{0.45\linewidth}
$\textbf{If } x_1 > \tau_1 \textbf{ then } \wt{a}_1'$; \\ 
$\textbf{else if } x_2 > \tau_{22} \textbf{ then } \wt{a}_2'$; \\ 
$\textbf{else if } x_1 \leq \tau_{32} \text{ and } x_2 > \tau_{33} \textbf{ then } \wt{a}_3$; \\ 
$\textbf{else } \wt{a}_3'$. 
\end{minipage}%
\hfill

\caption{ 
Diagram and description of the decision list $ \lb (\wt{c}_1', \wt{a}_1'), (\wt{c}_{2}', \wt{a}_{2}'), (\wt{c}_{3}, \wt{a}_{3}), \wt{a}_{3}' \rb $. 
Some of the values of $\wt{a}_1', \wt{a}_2', \wt{a}_3, \wt{a}_3'$ can be equal.
}
\label{fig:step6.1}
\end{figure}

\item The algorithm finishes Step~4, because $\Pi_\text{temp}$ contains no element now.

\item The algorithm proceeds to Step~5.

\begin{itemize}
\item We would like to pick a decision list from $\Pi_\text{final}$ that maximizes $\wh{R}(\cdot)$.

\item In this example, we have three decision lists in $\Pi_\text{final}$:
$\lb (\wt{c}_1, \wt{a}_1), \wt{a}_1' \rb$ with estimated value $15$,
$\lb (\wt{c}_1', \wt{a}_1'), (\wt{c}_{2}, \wt{a}_{2}), \wt{a}_{2}' \rb$ with estimated value $18$,
and
$\lb (\wt{c}_1', \wt{a}_1'), (\wt{c}_{2}', \wt{a}_{2}'), \wt{a}_{2} \rb$ with estimated value $18$.

\item We then choose the one with the maximal estimated value (with ties broken using the first encountered).
Therefore, the estimated optimal decision list $\wt{\pi}$ is described by
$ \lb (\wt{c}_1', \wt{a}_1'), (\wt{c}_{2}, \wt{a}_{2}), \wt{a}_{2}' \rb $,
as shown in Figure~\ref{fig:step4.1}.
\end{itemize}

\end{itemize}

\section[Asymptotic Properties of R(pi) for a Given pi]{Asymptotic Properties of $\wh{R}(\pi)$ for a Given $\pi$}

We shall derive some asymptotic properties of the doubly robust estimator $\wh{R}(\pi)$
introduced in Section~2.2 in the main paper.
In the next section, we will use these properties to 
derive an estimator for $\var \big\{ \wh{R}(\pi_1) - \wh{R}(\pi_2) \big\}$,
which is used by our proposed algorithm for finding an optimal decision list.

Hereafter denote the observed data for the $i$th subject by $O_i = (X_i^\T, A_i, Y_i)^\T$. 

% gamma
We first derive an i.i.d.\@ representation of 
$\wh\gamma = (\wh\gamma_1^\T, \dots, \wh\gamma_{m-1}^\T)^\T$,  
the maximum likelihood estimator of $\gamma = (\gamma_1^\T, \dots, \gamma_{m-1}^\T)^\T$
in the multinomial logistic regression model:
\[
\pr(A=a|X=x) =
\exp\left(u^{\T}\gamma_a\right) \bigg/ \left\lbrace
  1+\sum_{j=1}^{m-1}\exp\left(u^{\T}\gamma_j\right)
\right\rbrace.
\]
If $u = u(x) \equiv 1$, then the maximum likelihood estimator of 
$\omega(x, a) = \pr(A=a|X=x)$ reduces to $\E_{n}I(A=a)$. 
Thus the multinomial logistic regression model includes the sample proportion as its special case.
The log-likelihood function is
\begin{align*}
\loglikgamma(\gamma) &= \frac{1}{n} \sum_{i=1}^n \left[ \sum_{a=1}^{m-1} I(A_i = a) U_i^\T \gamma_a
  - \log\left\{ 1 + \sum_{a=1}^{m-1} \exp(U_i^\T \gamma_a) \right\} \right] \\
  &= \frac{1}{n} \sum_{i=1}^n \left[ \sum_{a=1}^{m-1} I(A_i = a) U_i^\T  \extgamma_a \gamma
  - \log\left\{ 1 + \sum_{a=1}^{m-1} \exp(U_i^\T \extgamma_a \gamma) \right\} \right],
\end{align*}
where $U_i = u(X_i)$, $q$ is the dimension of $U_i$, and
$\extgamma_1 = \big(I_q \mid \zero{q \times (m-2)q}\big),
\extgamma_2 = \big(\zero{q \times q} \mid I_q \mid \zero{q \times (m-3)q}\big), 
\dots,
\extgamma_{m-1} = \big(\zero{q \times (m-2)q} \mid I_q\big)$
are $(m-1)$ matrices of size $q \times (m-1)q$ satisfying $\extgamma_a \gamma = \gamma_a$.
Hence we have
\begin{align}
\frac{\partial \loglikgamma(\gamma)}{\partial \gamma} 
  &= \frac{1}{n} \sum_{i=1}^n \left\{ \sum_{a=1}^{m-1} I(A_i = a) \extgamma_a^\T U_i
  - \frac{ \sum_{a=1}^{m-1} \exp(U_i^\T \extgamma_a \gamma) \extgamma_a^\T U_i }
  { 1 + \sum_{a=1}^{m-1} \exp(U_i^\T \extgamma_a \gamma) }
  \right\}, \notag \\
\frac{\partial^2 \loglikgamma(\gamma)}{\partial \gamma \partial \gamma^\T}
  &= - \frac{1}{n} \sum_{i=1}^n 
  \frac{ \sum_{a=1}^{m-1} \exp(U_i^\T \extgamma_a \gamma) \extgamma_a^\T U_i U_i^\T \extgamma_a} 
  { 1 + \sum_{a=1}^{m-1} \exp(U_i^\T \extgamma_a \gamma) } \notag \\
  &\quad + \frac{1}{n} \sum_{i=1}^n 
  \frac{ \{ \sum_{a=1}^{m-1} \exp(U_i^\T \extgamma_a \gamma) \extgamma_a^\T U_i \} 
    \{ \sum_{a=1}^{m-1} \exp(U_i^\T \extgamma_a \gamma) U_i^\T \extgamma_a \} }
  { \{ 1 + \sum_{a=1}^{m-1} \exp(U_i^\T \extgamma_a \gamma) \}^2 }.
\label{eq:loglikgamma-deriv}
\end{align}
Denote $\gamma_0$ as the maximizer of $\E \loglikgamma(\gamma)$.
By the likelihood theory, we have
\[
\sqrt{n} (\wh\gamma - \gamma_0) =
  - \sqrt{n} \left[ \E \left\{ \frac{\partial^2 \loglikgamma(\gamma_0)}{\partial \gamma \partial \gamma^\T} \right\} \right]^{-1}
  \left\{ \frac{\partial \loglikgamma(\gamma_0)}{\partial \gamma} \right\} + o_p(1),
\]
where the partial derivatives are given in \eqref{eq:loglikgamma-deriv}, 
and $o_p(1)$ denotes a random quantity that convergences to zero in probability.
Define
\[
\ifgamma(O_i) = 
  - \left\{ \E \left( \frac{\partial^2 \loglikgamma(\gamma_0)}{\partial \gamma \partial \gamma^\T} \right) \right\}^{-1}
  \left\{ \sum_{a=1}^{m-1} I(A_i = a) \extgamma_a^\T U_i
  - \frac{ \sum_{a=1}^{m-1} \exp(U_i^\T \extgamma_a \gamma_0) \extgamma_a^\T U_i }
  { 1 + \sum_{a=1}^{m-1} \exp(U_i^\T \extgamma_a \gamma_0) }
  \right\}.
\]
Then we have
\[
\sqrt{n} (\wh\gamma - \gamma_0) = \frac{1}{\sqrt{n}} \sum_{i=1}^n \ifgamma(O_i) + o_p(1).
\]

% beta
Next we derive an i.i.d.\@ representation of 
$\wh\beta = (\wh\beta_1^\T, \dots, \wh\beta_m^\T)^\T$,  
the maximum likelihood estimator of $\beta = (\beta_1^\T, \dots, \beta_m^\T)^\T$
in the generalized linear model:
\[
g\left\{ \E(Y_i | X_i, A_i) \right\} = \sum_{a=1}^m I(A_i = a) Z_i^\T \beta_a.
\]
We assume that $Y_i$ given $A_i$ and $X_i$ has an distribution in the exponential family with density function
\[
f_{Y_i}(y_i) = \exp \left\{ \frac{y_i \theta_i - b(\theta_i)}{\phi} + c(y_i, \phi) \right\},
\]
where $\theta_i$ and $\phi$ are parameters, 
and $b(\cdot)$ and $c(\cdot, \cdot)$ are known functions.
Note that for normal distribution $\phi$ is known as the dispersion parameter while
for Bernoulli distribution $\phi$ is always equal to one.
For simplicity we assume $g(\cdot)$ is a canonical link function hereafter. 
Then we have $b'(\cdot) \equiv g^{-1}(\cdot)$ and
$\theta_i = \sum_{a=1}^m I(A_i = a) Z_i^\T \beta_a$.
The log-likelihood function is
\begin{align*}
\loglikbeta(\beta, \phi)
  &= \frac{1}{n} \sum_{i=1}^n
  \left[ \frac{ Y_i \sum_{a=1}^m I(A_i = a) Z_i^\T \beta_a 
    - b \left\{ \sum_{a=1}^m I(A_i = a) Z_i^\T \beta_a \right\} }
  {\phi} + c(Y_i, \phi) \right] \\
  &= \frac{1}{n} \sum_{i=1}^n
  \left[ \frac{ Y_i \sum_{a=1}^m I(A_i = a) Z_i^\T \extbeta_a \beta 
    - b \left\{ \sum_{a=1}^m I(A_i = a) Z_i^\T \extbeta_a \beta \right\} }
  {\phi} + c(Y_i, \phi) \right],
\end{align*}
where $r$ is the dimension of $Z_i$, and 
$\extbeta_1 = \big(I_q \mid \zero{q \times (m-1)q}\big),
\extbeta_2 = \big(\zero{q \times q} \mid I_q \mid \zero{q \times (m-2)q}\big), 
\dots,
\extbeta_m \break = \big(\zero{q \times (m-1)q} \mid I_q\big)$
are $m$ matrices of size $q \times mq$ satisfying $\extbeta_a \beta = \beta_a$.
Then we have
\begin{align*}
\frac{\partial \loglikbeta(\beta, \phi)}{\partial \beta} &=
  \frac{1}{n \phi} \sum_{i=1}^n 
  \left[ Y_i - b' \left\{ \sum_{a=1}^m I(A_i = a) Z_i^\T \extbeta_a \beta \right\} \right]
  \left\{  \sum_{a=1}^m I(A_i = a) \extbeta_a^\T Z_i \right\}, \\
\frac{\partial^2 \loglikbeta(\beta, \phi)}{\partial \beta \partial \beta^\T} &=
  - \frac{1}{n \phi} \sum_{i=1}^n 
  b'' \left\{ \sum_{a=1}^m I(A_i = a) Z_i^\T \extbeta_a \beta \right\}
  \left\{  \sum_{a=1}^m I(A_i = a) \extbeta_a^\T Z_i Z_i^\T \extbeta_a \right\}.
\end{align*}
By the property of the score function, we have
\[
\E \left( \frac{\partial^2 \loglikbeta(\beta_0, \phi_0)}{\partial \beta \partial \phi} \right)
= -\frac{1}{\phi} \E \left( \frac{\partial \loglikbeta(\beta_0, \phi_0)}{\partial \beta} \right) = 0.
\]
Therefore, by the likelihood theory and the property of block diagonal matrix, we conclude that
\begin{align*}
\sqrt{n} (\wh\beta - \beta_0) &= 
  - \sqrt{n} \left[ \E \left\{ \frac{\partial^2 \loglikbeta(\beta_0, \phi_0)}{\partial \beta \partial \beta^\T} \right\} \right]^{-1}
  \left\{ \frac{\partial \loglikbeta(\beta_0, \phi_0)}{\partial \beta} \right\}
  + o_p(1) \\
  &= \frac{1}{\sqrt{n}} \sum_{i=1}^n \ifbeta(O_i) + o_p(1),
\end{align*}
where
\begin{multline*}
\ifbeta(O_i) = \left( \E \left[
  b'' \left( \sum_{a=1}^m I(A_i = a) Z_i^\T \extbeta_a \beta_0 \right)
  \left\{  \sum_{a=1}^m I(A_i = a) \extbeta_a^\T Z_i Z_i^\T \extbeta_a \right\}
  \right] \right)^{-1} \\
  \cdot 
  \left\{ Y_i - b' \left( \sum_{a=1}^m I(A_i = a) Z_i^\T \extbeta_a \beta_0 \right) \right\}
  \left\{  \sum_{a=1}^m I(A_i = a) \extbeta_a^\T Z_i \right\}.
\end{multline*}

% R(\pi)
Finally we derive an i.i.d.\@ representation of $\wh{R}(\pi)$. 
To emphasize the dependence of $\omega(x, a)$ and $\mu(x, a)$ on the parameters $\gamma$ and $\beta$,
in the following we write $\omega(x, a)$ as $\omega(x, a, \gamma)$
and $\mu(x, a)$ as $\mu(x, a, \beta)$.
Thus we have
$\wh\omega(x, a) = \omega(x, a, \wh\gamma)$
and $\wh\mu(x, a) = \mu(x, a, \wh\beta)$.
Note that
\begin{align*}
\omega(x, a, \gamma) &= \frac{\exp(u^\T \extgamma_a \gamma)}
  {\sum_{j=1}^m \exp(u^\T \extgamma_j \gamma)}, \\
\mu(x, a, \beta) &= b'(z^\T \extbeta_a \beta), 
\end{align*}
for $a = 1, \dots, m$, where $\extgamma_m = \zero{q \times (m-1)q}$.
Hence we have
\begin{align}
\frac{\partial \omega(x, a, \gamma)}{\partial \gamma}
  &= \frac{ \exp(u^\T \extgamma_a \gamma) 
    \left\{ \sum_{j=1}^m \exp(u^\T \extgamma_j \gamma)  \cdot
    (\extgamma_a^\T - \extgamma_j^\T) u \right\} }
    { \left\{ \sum_{j=1}^m \exp(u^\T \extgamma_j \gamma) \right\}^2 }, \notag \\
\frac{\partial \mu(x, a, \beta)}{\partial \beta}
  &= b''(z^\T \extbeta_a \beta) \extbeta_a^\T z.
\label{eq:deriv}
\end{align}
By Taylor expansion, we have
\begin{align*}
\wh{R}(\pi) 
&= \frac{1}{n} \sum_{i=1}^n \sum_{a=1}^{m} 
  \left[ \frac{I(A_i = a)}{\omega(X_i, a, \wh\gamma)} 
  \left\{Y_i - \mu(X_i, a, \wh\beta) \right\}  + \mu(X_i, a, \wh\beta) \right] I\{\pi(X_i) = a\} \\
&= \frac{1}{n} \sum_{i=1}^n \sum_{a=1}^{m} 
  \left[ \frac{I(A_i = a)}{\omega(X_i, a, \gamma_0)} 
  \left\{Y_i - \mu(X_i, a, \beta_0) \right\}  + \mu(X_i, a, \beta_0) \right] I\{\pi(X_i) = a\} \\
& \quad + \frac{1}{n} \sum_{i=1}^n \sum_{a=1}^{m} 
  \left[ -\frac{I(A_i = a)}{\omega^2(X_i, a, \gamma_0)} 
  \left\{Y_i - \mu(X_i, a, \beta_0) \right\} I\{\pi(X_i) = a\}  
  \frac{\partial \omega(X_i, a, \gamma_0)}{\partial \gamma} \right]^\T (\wh\gamma - \gamma_0) \\
& \quad + \frac{1}{n} \sum_{i=1}^n \sum_{a=1}^{m} 
  \left[ \left\{ -\frac{I(A_i = a)}{\omega(X_i, a, \gamma_0)} + 1 \right\} I\{\pi(X_i) = a\}
  \frac{\partial \mu(X_i, a, \beta_0)}{\partial \beta} \right]^\T (\wh\beta - \beta_0)
  + o_p(1) \\
&= \frac{1}{n} \sum_{i=1}^n \sum_{a=1}^{m} 
  \left[ \frac{I(A_i = a)}{\omega(X_i, a, \gamma_0)} 
  \left\{Y_i - \mu(X_i, a, \beta_0) \right\}  + \mu(X_i, a, \beta_0) \right] I\{\pi(X_i) = a\} \\
& \quad + \E \left( \sum_{a=1}^{m} 
  \left[ -\frac{I(A_i = a)}{\omega^2(X_i, a, \gamma_0)} 
  \left\{Y_i - \mu(X_i, a, \beta_0) \right\} I\{\pi(X_i) = a\} 
  \frac{\partial \omega(X_i, a, \gamma_0)}{\partial \gamma} \right] \right)^\T
   (\wh\gamma - \gamma_0) \\
& \quad + \E \left( \sum_{a=1}^{m} 
  \left[ \left\{ -\frac{I(A_i = a)}{\omega(X_i, a, \gamma_0)} + 1 \right\} I\{\pi(X_i) = a\} 
  \frac{\partial \mu(X_i, a, \beta_0)}{\partial \beta} \right] \right)^\T
  (\wh\beta - \beta_0)
  + o_p(1).
\end{align*}
Recall that
\[
R(\pi) = \E \left( \sum_{a=1}^m \left[ \frac{I(A_i = a)}{\omega(X_i, a, \gamma_0)} 
  \left\{Y_i - \mu(X_i, a, \beta_0) \right\}  + \mu(X_i, a, \beta_0) \right] I\{\pi(X_i) = a\} \right).
\]
Define
\begin{align}
\ifR(O_i) &= \sum_{a=1}^m \left[ \frac{I(A_i = a)}{\omega(X_i, a, \gamma_0)} 
  \left\{Y_i - \mu(X_i, a, \beta_0) \right\}  + \mu(X_i, a, \beta_0) \right] I\{\pi(X_i) = a\} \notag \\
  & \quad - \E \left( \sum_{a=1}^m \left[ \frac{I(A_i = a)}{\omega(X_i, a, \gamma_0)} 
  \left\{Y_i - \mu(X_i, a, \beta_0) \right\}  + \mu(X_i, a, \beta_0) \right] I\{\pi(X_i) = a\} \right) \notag \\
  & \quad + \E \left( \sum_{a=1}^{m} 
  \left[ -\frac{I(A_i = a)}{\omega^2(X_i, a, \gamma_0)} 
  \left\{Y_i - \mu(X_i, a, \beta_0) \right\} I\{\pi(X_i) = a\} 
  \frac{\partial \omega(X_i, a, \gamma_0)}{\partial \gamma} \right] \right)^\T \ifgamma(O_i) \notag \\ 
  & \quad + \E \left( \sum_{a=1}^{m} 
  \left[ \left\{ -\frac{I(A_i = a)}{\omega(X_i, a, \gamma_0)} + 1 \right\} I\{\pi(X_i) = a\} 
  \frac{\partial \mu(X_i, a, \beta_0)}{\partial \beta} \right] \right)^\T \ifbeta(O_i),
\label{eq:ifR}
\end{align}
where $\partial \omega / \partial \gamma$ and $\partial \mu / \partial \beta$
are given in \eqref{eq:deriv}.
Then we have
\[
\sqrt{n} \left\{ \wh{R}(\pi) - R(\pi) \right\} = \frac{1}{\sqrt{n}} \sum_{i=1}^n \ifR(O_i) + o_p(1).
\]
Therefore, by the central limit theorem and the Slutsky's theorem, we conclude that
\begin{equation}
\label{eq:R-AN}
\sqrt{n} \left\{ \wh{R}(\pi) - R(\pi) \right\} \stackrel{d}{\to} N(0, \E \left\{ \ifR^2(O_i) \right\}),
\end{equation}
where $\stackrel{d}{\to}$ denotes convergence in distribution.

To estimate the asymptotic variance, we use the plug-in method.
Namely, define $\ifRhat(O_i)$ as in \eqref{eq:ifR} except that expectations are replaced with sample averages
and true values are replaced with corresponding estimates.
Then $\var\left(\wh{R}(\pi)\right)$ can be estimated by $\sum_{i=1}^n \ifRhat^2(O_i) / n^2$.

\section[Asymptotic Properties of R(pi1) - R(pi2)]{Asymptotic Properties of $\wh{R}(\pi_1) - \wh{R}(\pi_2)$}

Define $\ifRa(O_i)$ as in \eqref{eq:ifR} with $\pi$ replaced by $\pi_1$. 
Define $\ifRb(O_i)$ as in \eqref{eq:ifR} with $\pi$ replaced by $\pi_2$.
Define $\ifRahat(O_i)$ and $\ifRbhat(O_i)$ similarly.
Then we have
\begin{align*}
\sqrt{n} \left[ \left\{ \wh{R}(\pi_1) - \wh{R}(\pi_2) \right\} - \left\{ R(\pi_1) - R(\pi_2) \right\} \right] 
  &= \frac{1}{\sqrt{n}} \sum_{i=1}^n \left\{ \ifRa(O_i) - \ifRb(O_i) \right\} + o_p(1) \\
  &\stackrel{d}{\to} N(0, \E \left\{ \ifRa(O_i) - \ifRb(O_i) \right\}^2).
\end{align*}
Therefore, we can estimate $\var \left\{ \wh{R}(\pi_1) - \wh{R}(\pi_2) \right\}$ by
\begin{equation}
\label{eq:var-delta}
\wh{\var} \left\{ \wh{R}(\pi_1) - \wh{R}(\pi_2) \right\} = \frac{1}{n^2} \sum_{i=1}^n \left\{ \ifRahat(O_i) - \ifRbhat(O_i) \right\}^2.
\end{equation}

The variance estimator $\wh{\var}(\Delta_j)$ used in the algorithm in Section~2.4.1 in the main paper
can be obtained via \eqref{eq:var-delta}
with
$\pi_1 = \lb (\wb{c}_1, \wb{a}_1), \dots, (\wb{c}_{j-1}, \wb{a}_{j-1}), (\wt{c}_{j}, \wt{a}_{j}), \wt{a}_{j}' \rb$
and
$\pi_2 = \lb (\wb{c}_1, \wb{a}_1), \dots, (\wb{c}_{j-1}, \wb{a}_{j-1}), \wb{a}_{j-1}' \rb$.

\section{Implementation Details of Finding an Optimal Decision List}

\subsection{Algorithm Description}

We give an equivalent version of the proposed algorithm
for finding an optimal decision list.
Compared to the algorithm presented in the main paper,
this version makes use of recursive calls to avoid explicit constructions of 
sets $\Pi_\text{temp}$ and $\Pi_\text{final}$,
and facilitates the analysis of time complexity.
The algorithm is as follows.

\vspace*{1ex}
\begin{algorithm}
\SetKwInput{KwIn}{Input}
\SetKwInput{KwOut}{Output}
\SetKwFunction{FindList}{FindList}
\KwIn{$\wh{R}(\cdot)$, $L_\text{max}$, $\alpha$}
\KwOut{a decision list $\wt\pi$ that maximize $\wh{R}(\cdot)$}

$ \wt{a}_{0} = \argmax_{a_0\in\mathcal{A}}
  \wh{R}\left[ \lb a_0 \rb \right] $\;
$ \wt{\pi} = \FindList(1, \{ \}, \wt{a}_{0}) $\;

\end{algorithm}
\vspace*{1ex}

The function \FindList is defined below. When $j = 1$, we treat 
$ (\wb{c}_1, \wb{a}_1),\dots, (\wb{c}_{j-1}, \wb{a}_{j-1}) $
as an empty array.
Thus when $j = 1$,
$ \lb (\wb{c}_1, \wb{a}_1),\dots, (\wb{c}_{j-1}, \wb{a}_{j-1}), (\wt{c}_j, \wt{a}_j), \wt{a}_j' \rb $
is the same as $ \lb (\wt{c}_1, \wt{a}_1), \wt{a}_1' \rb $ and 
$ \lb (\wb{c}_1, \wb{a}_1),\dots, (\wb{c}_{j-1}, \wb{a}_{j-1}), \wb{a}_{j-1}' \rb $
is the same as $ \lb \wb{a}_0' \rb $.

\vspace*{1ex}
\begin{algorithm}
\SetKwProg{Fn}{Function}{}{end}

\Fn{\FindList($j,\; \lb (\wb{c}_1, \wb{a}_1), \dots, (\wb{c}_{j-1}, \wb{a}_{j-1}) \rb,\; \wb{a}_{j-1}'$)}
{
  $ (\wt{c}_{j}, \wt{a}_{j}, \wt{a}_{j}') = 
    \argmax_{(c_{j}, a_{j}, a_{j}') \in \mathcal{C} \times \mathcal{A} \times \mathcal{A}} 
    \wh{R}\left[ \lb (\wb{c}_1, \wb{a}_1), \dots, (\wb{c}_{j-1}, \wb{a}_{j-1}), (c_j, a_j), a_j' \rb \right] $\;
  $ \wh{\Delta}_j = 
    \wh{R}\left[ \lb (\wb{c}_1, \wb{a}_1), \dots, (\wb{c}_{j-1}, \wb{a}_{j-1}), (\wt{c}_j, \wt{a}_j), \wt{a}_j' \rb \right] 
    - \wh{R}\left[ \lb (\wb{c}_1, \wb{a}_1), \dots, (\wb{c}_{j-1}, \wb{a}_{j-1}), \wb{a}_{j-1}' \rb \right] $\;

  \uIf{$\wh{\Delta}_j < z_{1-\alpha} \big\{ \wh{\mathrm{Var}}\big(\wh{\Delta}_j\big) \big\}^{1/2}$} {
    $\wt{\pi} = \lb (\wb{c}_1, \wb{a}_1), \dots, (\wb{c}_{j-1}, \wb{a}_{j-1}), \wb{a}_{j-1}' \rb$\;
  }
  \uElseIf{$j = L_\mathrm{max}$}
  {
    $\wt{\pi} = \lb (\wb{c}_1, \wb{a}_1), \dots, (\wb{c}_{j-1}, \wb{a}_{j-1}), (\wt{c}_j, \wt{a}_j), \wt{a}_j' \rb$\;
  }
  \Else{
    $\wt{\pi}_1 = \FindList(j+1,\; \lb (\wb{c}_1, \wb{a}_1), \dots, (\wb{c}_{j-1}, \wb{a}_{j-1}), (\wt{c}_j, \wt{a}_j) \rb,\; \wt{a}_j')$\;
    $\wt{\pi}_2 = \FindList(j+1,\; \lb (\wb{c}_1, \wb{a}_1), \dots, (\wb{c}_{j-1}, \wb{a}_{j-1}), (\wt{c}_j', \wt{a}_j') \rb,\; \wt{a}_j)$, 
      \linebreak\hspace*{2em}
      where $\wt{c}_j' = \text{negation of}\ \wt{c}_j$\;
    $\wt{\pi} = \argmax_{\pi \in \{ \wt{\pi}_1, \wt{\pi}_2 \}} \wh{R}(\pi)$\;
  }
  \Return $\wt{\pi}$\;
}

\end{algorithm}
\vspace*{1ex}

In the \FindList function, a crucial step is to compute 
$ (\wt{c}_{j}, \wt{a}_{j}, \wt{a}_{j}') $.
A straightforward implementation that involves a brute-force search over 
$\mathcal{C} \times \mathcal{A} \times \mathcal{A}$
can be time consuming.
We provide an efficient implementation below.

We observe that some calculations can be performed only once at the beginning of the algorithm. 
First, define 
\[
\wh{\xi}_{ia} = 
  \frac{I(A_i = a)}{\omega(X_i, a, \wh\gamma)} 
  \left\{Y_i - \mu(X_i, a, \wh\beta) \right\}  + \mu(X_i, a, \wh\beta).
\]
Then we have
\[
\wh{R}(\pi) 
= \frac{1}{n} \sum_{i=1}^n \sum_{a=1}^{m} 
  \wh{\xi}_{ia} I\{\pi(X_i) = a\}.
\]
Second, for the $i$th subject, denote $x_{ij}$ as the observed value of his/her $j$th covariate.
For the $j$th baseline covariate, 
there are $s_k = \#\mathcal{X}_j$ possible candidate cutoff values $\tau_{j 1} \leq \dots \leq \tau_{j s_j}$, 
which divides the real line into $s_k + 1$ intervals:
\[
(-\infty, \tau_{j 1}] ,\, (\tau_{j 1}, \tau_{j 2}],\, \dots,\, 
(\tau_{j (s_j - 1)}, \tau_{j s_j}],\, (\tau_{j s_j}, \infty).
\]
Then we code the observed values $x_{1j}, \dots, x_{nj}$ into indices 
$b_{1j}, \dots, b_{nj}$ according to which interval they fall.

In order to reduce the number of evaluations of $\wh{R}(\cdot)$ when searching for the maximizer over 
$\mathcal{C} \times \mathcal{A} \times \mathcal{A}$,
we organize the intermediate results as shown below.
Let $\mathcal{I} = \{i: X_i \in \mathcal{T}(\wb{c}_\ell)^c \text{ for all } \ell < j\}$.
Then $\mathcal{I}$ contains all the subjects that have not had treatment recommendations up to the $j$th clause.
Since we have
\[
n \wh{R}(\pi)
= \sum_{i \in \mathcal{I}} \sum_{a=1}^{m} \wh{\xi}_{ia} I\{\pi(X_i) = a\} 
+ \sum_{i \in \mathcal{I}^c} \sum_{a=1}^{m} \wh{\xi}_{ia} I\{\pi(X_i) = a\}
\]
and $\sum_{i \in \mathcal{I}^c} \sum_{a=1}^{m} \wh{\xi}_{ia} I\{\pi(X_i) = a\}$ is constant during the maximization, 
we focus on maximizing 
$\sum_{i \in \mathcal{I}} \sum_{a=1}^{m} \wh{\xi}_{ia} I\{\pi(X_i) = a\}$, 
which reduces to maximizing 
\begin{equation}
\label{eq:efficient-implementation}
\sum_{i \in \mathcal{I}} \sum_{a=1}^{m} \wh{\xi}_{ia} I\{ i \in \mathcal{T}(c_{j}), a = a_{j} \}
+ \sum_{i \in \mathcal{I}} \sum_{a=1}^{m} \wh{\xi}_{ia} I\{ i \notin \mathcal{T}(c_{j}), a = a_{j}' \}.
\end{equation}
To identify the maximizer of \eqref{eq:efficient-implementation}, 
we first loop over all possible pairs of covariates.
For each pair of covariates, say the $k$th and the $\ell$th covariates, 
define $D$, a three-dimensional array of size $m \times (s_k + 1) \times (s_\ell + 1)$,
as 
$D_{auv} = \sum_{i \in \mathcal{I}} \wh{\xi}_{ia} I(b_{ik} = u, b_{il} = v)$.
Next, we loop over all possible cutoff values
and construct the corresponding $c_{j}$.
The values of $a_{j}$ and $a_{j}'$ that maximizes \eqref{eq:efficient-implementation} for a given $c_{j}$
can be easily obtained due to the additive structure.
After enumerating all the possible conditions that $c_{j}$ may take,
we can find out $ (\wt{c}_{j}, \wt{a}_{j}, \wt{a}_{j}') $.

\subsection{Time Complexity Analysis}

Since computing $\wh{\xi}_{ia}$s requires $O(nm)$ time and 
computing $b_{ij}$s requires $O(np)$ time.
The calculations at the beginning of the algorithm take $O(nm + np)$ time in total.

The algorithm first computes $\wt{a}_{0}$,
which requires $O(nm)$ time.
Then it invokes a function call $ \FindList(1, \{ \}, \wt{a}_{0}) $.
Due to the recursive nature of the \FindList function, we will compute the time complexity
by establishing a recurrence relation between $T(j)$ and $T(j+1)$,
where $T(j)$ is the time complexity of the function call 
\FindList($j$, $\left\lbrace (\wb{c}_1, \wb{a}_1),\dots, (\wb{c}_j, \wb{a}_j) \right\rbrace$).

Suppose a call \FindList($j$, $\left\lbrace (\wb{c}_1, \wb{a}_1),\dots, (\wb{c}_j, \wb{a}_j) \right\rbrace$) is invoked. 
The running time can be computed by going through the algorithm of the \FindList function step-by-step as follows.

First, the function computes $ (\wt{c}_{j}, \wt{a}_{j}, \wt{a}_{j}')$.
A naive implementation would involve looping over all the covariates, all the possible cutoff values and all the treatment options, whose running time is $O(n m p^2 s^2)$, where $s = \max_j s_j$.
However, the running time is greatly reduced if we use the efficient implementation described previously.
For a given pair of covariates,
we can compute $D_{auv}$s in $O(nm)$ time.
Then we can find out the maximum of \eqref{eq:efficient-implementation} in $O(m s^2)$ time
by looping over all possible cutoff values.
Therefore, the total time for computing $(\wt{c}_{j}, \wt{a}_{j}, \wt{a}_{j}')$
is $O\{ (n + s^2) m p^2 \}$.

Second, the function computes $ \wh{\Delta}_j $, which takes $O(n)$ time.

Third, the function computes $\wh{\var}\big( \wh{\Delta}_j \big)$,
whose running time is $O(nmq + nmr)$,
where $q$ is the dimension of $U_i$ and $r$ is the dimension of $Z_i$.
Since both $U_i$ and $Z_i$ are known feature vectors constructed from $X_i$,
for most cases $q$ and $r$ are of the same order as $p$.
So this step takes $O(nmp)$ time.

Fourth, the function executes the ``if-then'' statement.
In the worst case, the function makes two recursive calls, taking $2 T(j+1)$ time.

Combining these four steps, we have
$ T(j) = O\{ (n + s^2) m p^2 \} + 2 T(j + 1) $.
The boundary condition is $T(L_\mathrm{max}) = O\{ (n + s^2) m p^2 \}$.
Using backward induction, we get $T(0) = O\{ 2^{L_\text{max}} (n + s^2) m p^2 \}$.
Recall that $s = \max_j \#\mathcal{X}_j$.

Combining $T(0)$ with the running time before invoking $ \FindList(1, \{ \}, \wt{a}_{0}) $, 
we obtain that the time complexity of the entire algorithm is 
$O[ 2^{L_\text{max}} m p^2 \{n + (\max_j \#\mathcal{X}_j)^2\} ]$.

\section{Implementation Details of Finding an Equivalent Decision List with Minimal Cost}

In this section we give an algorithmic description of the proposed method
for finding an equivalent decision list with minimal cost.
Recall that two decision lists are called equivalent if they give the same treatment recommendation 
for every patient in the population.

\vspace*{1ex}
\begin{algorithm}
\SetKwInput{KwIn}{Input}
\SetKwInput{KwOut}{Output}
\SetKwFunction{FindMinCost}{FindMinCost}

\KwIn{a decision list $\bar{\pi}$}
\KwOut{an equivalent decision list $\pi_\mathrm{min}$ with minimal cost $N_\mathrm{min}$}
Identify atoms in $\bar{\pi}$ as $d_1, \dots, d_K$\;
Compute $\mathcal{I}_a = \{i: \bar{\pi}(X_i) = a\}$ for each $a \in \mathcal{A}$\;
Set $\pi_\mathrm{min} = \{ \}$ and $N_\mathrm{min} = \infty$\;
\FindMinCost($0$, $ \{ \} $, $\pi_\mathrm{min}$, $N_\mathrm{min}$)\;
\end{algorithm}
\vspace*{1ex}

The function \FindMinCost is defined below.
\pagebreak

\vspace*{1ex}
\begin{algorithm}
\SetKwProg{Fn}{Function}{}{end}
\Fn{\FindMinCost($j$, $\{(c_1, a_1), \dots, (c_j, a_j)\}$, $\pi_{\text{min}}$, $N_{\text{min}}$)}{
  Compute a lower bound of the cost as
  $
  N_\text{bd} = \mathcal{N}_\ell \sum_{\ell=1}^j \pr_n(X \in \mathcal{R}_\ell) \linebreak
    \hspace*{1.8em} + \mathcal{N}_j \pr_n(X \in \cap_{\ell=1}^j \mathcal{R}_\ell^c) 
  $,
  where $\pr_n$ denotes the empirical probability measure\;
  \lIf{$N_{\mathrm{bd}} \geq N_{\mathrm{min}}$}{\Return}

  $\mathcal{I} = \{i: X_i \in \mathcal{T}(c_\ell)^c \text{ for all } \ell \leq j\}$\;
  \eIf{$\mathcal{I} \subset \mathcal{I}_{a_0}$ for some $a_0$}{
    \If{$N[ \{(c_1, a_1), \dots, (c_j, a_j), a_0\} ] < N_\mathrm{min}$}{
      $\pi_\text{min} = \{(c_1, a_1), \dots, (c_j, a_j), a_0\}$\;
      $N_\text{min} = N(\pi_\text{min})$\;
    }
  }{
    \For{$1 \leq k_1 < k_2 \leq K$}{
      Let $\mathcal{C}_{k_1, k_2}$ be the set consisting of all the logical clauses involving \linebreak\hspace*{1.8em}
        $d_{k_1}$ or $d_{k_2}$ or both using conjunction, disjunction, and/or negation\;
      \For{$c_{j+1} \in \mathcal{C}_{k_1, k_2}$}{
        $\mathcal{J}_{j+1} = \{i \in \mathcal{I} : X_i \in \mathcal{T}(c_{j+1}) \}$\;
        \lIf{$\mathcal{J}_{j+1}$ is non-empty and $\mathcal{J}_{j+1} \subset \mathcal{I}_{a_{j+1}}$ for some $a_{j+1} \in \mathcal{A}$}{
          \hspace*{1.8em}\FindMinCost($j+1$, $\{(c_1, a_1), \dots, (c_j, a_j), (c_{j+1}, a_{j+1})\}$, $\pi_{\text{min}}$, $N_{\text{min}}$)
        }
      }
    }
  }
}
\end{algorithm}
\vspace*{1ex}

\section[Point Estimate and Prediction Interval for R(pi) with Bootstrap Bias Correction]{Point Estimate and Prediction Interval for $R(\wh{\pi})$ with Bootstrap Bias Correction}

In this section, we show how to estimate the value of the estimated treatment regime, $R(\wh{\pi})$,
and how to construct a prediction interval for it.

\subsection{Methodology}

To measure how well the estimated
treatment regime $\wh{\pi}$ performs, it is often of
interest to construct an estimator of and a prediction interval for
$R(\wh{\pi})$. 
Since a natural candidate for estimating $R(\wh{\pi})$ is $\wh{R}(\wh{\pi})$,
it may be tempting to construct a prediction interval centering at $\wh{R}(\wh{\pi})$.
However, $\wh{R}(\wh{\pi})$ is generally too optimistic to serve as an
honest estimator of $R(\wh{\pi})$. It has an upward
bias due to the maximization process.
As a remedy, we suggest using $B$ bootstraps to correct this bias.
Specifically, the perturbed version of $\wh{R}(\pi)$ in the $b$th bootstrapping sample is
\[
\wh{R}^*_b(\pi) = \frac{1}{n} \sum_{i=1}^n \left( W_i \sum_{a=1}^{m} 
  \left[ \frac{I(A_i = a)}{\omega(X_i, a, \wh\gamma^*)} 
  \left\{Y_i - \mu(X_i, a, \wh\beta^*) \right\}  + \mu(X_i, a, \wh\beta^*) \right] I\{\pi(X_i) = a\} \right),
\]
where $W_1, \dots, W_n$ are identically and independently distributed with standard exponential distribution,
$\wh\gamma^*$ is the solution to
\[
\sum_{i=1}^n W_i \left\{ \sum_{a=1}^{m-1} I(A_i = a) \extgamma_a^\T U_i
  - \frac{ \sum_{a=1}^{m-1} \exp(U_i^\T \extgamma_a \gamma) \extgamma_a^\T U_i }
  { 1 + \sum_{a=1}^{m-1} \exp(U_i^\T \extgamma_a \gamma) }
  \right\} = 0,
\]
and $\wh\beta^*$ is the solution to
\[
\sum_{i=1}^n W_i
  \left[ Y_i - b' \left\{ \sum_{a=1}^m I(A_i = a) Z_i^\T \extbeta_a \beta \right\} \right]
  \left\{  \sum_{a=1}^m I(A_i = a) \extbeta_a^\T Z_i \right\} = 0.
\]
Let $\wh{\pi}^*_b$ be the maximizer of $\wh{R}^*_b(\pi)$ 
over $\Pi$. 
Then the actual bias $\wh{R}(\wh{\pi}) - R(\wh{\pi})$ can be estimated by the corresponding bias in
the bootstrap world: $\wh{\bias} = \sum_{b=1}^B 
\{ \wh{R}^*_b(\wh{\pi}^*_{b}) - \wh{R}(\wh{\pi}^*_{b}) \} / B$,
where $B$ is the number of bootstrap samples.
The final estimator of $R(\wh{\pi})$ is 
$\wh{R}_\text{c}(\wh{\pi}) = \wh{R}(\wh{\pi}) - \wh{\bias}$.

To construct a prediction interval for $R(\wh{\pi})$,
we treat $\wh{\pi}$ as a non-random
quantity, and then utilize the asymptotic normality of $\wh{R}(\wh{\pi})$ given in \eqref{eq:R-AN}.
Let $z_\rho$ be the $100\rho$ percentile of a standard normal distribution and
$\wh{\sigma}^2 = \wh{\var} \big\{ \wh{R}(\wh{\pi}) \big\}$.
Then a $(1-\alpha) \times 100\%$ prediction interval for $R(\wh{\pi})$ is
\begin{equation}
\label{eq:ci}
\big[ \wh{R}_\text{c}(\wh{\pi}) + z_{\alpha/2} \wh{\sigma}, \;
   \wh{R}_\text{c}(\wh{\pi}) + z_{1-\alpha/2} \wh{\sigma} \big].
\end{equation}
A potential drawback of this interval is, though, that it ignores the variation
introduced by $\wh{\pi}$.
Nevertheless, our numerical experiences suggest that
this extra variation is generally small and the coverage probability 
is close to the nominal level.
Taking into account the variability of $\wh{\pi}$
has to deal with the associated non-regularity issue, 
which is beyond the scope of this paper.

As a final remark, for binary outcome 
we suggest to conduct the bias correction and construct the prediction interval
based on $\logit\{\wh{R}(\cdot)\}$ first and then transform back to the original scale,
where $\logit(v) = \log\{v / (1 - v)\}$.

\subsection{Simulations}

We present the point estimate and the coverage probabilities of the plain prediction interval and the prediction interval with bootstrap bias correction
in Table~\ref{tbl:ci}. 
The setting used here is exactly the same as that in Section~3 in the main paper.
We can see that the bias correction improves the coverage probability substantially in finite samples,
especially as the number of covariates gets larger.
Besides, the bootstrap prediction interval is prone to overcoverage for the binary response.

\begin{table}
\renewcommand{\baselinestretch}{1.1}\fontsize{11}{12}\selectfont
\centering
\caption{Point estimate and coverage probabilities of prediction intervals with and without bootstrap bias correction.
Plain-PI refers to the coverage probability of the plain prediction interval, and
Corrected-PI refers to the coverage probability of the bias-corrected prediction interval.}
\label{tbl:ci}
\begin{tabular*}{\columnwidth}{c *{9}{@{\extracolsep{\fill}} c}}
  \toprule
 \multirow{2}{*}{\;$p$\;} & \multirow{2}{*}{Setting} & 
  \multicolumn{4}{c}{Continuous response} & \multicolumn{4}{c}{Binary response} \\
  \cmidrule{3-6} \cmidrule{7-10}
 & & $R(\widehat{\pi})$ & $\widehat{R}_\text{c}(\widehat{\pi})$ & Plain-PI & Corrected-PI 
    & $R(\widehat{\pi})$ & $\widehat{R}_\text{c}(\widehat{\pi})$ & Plain-PI & Corrected-PI \\ 
  \midrule
  \multirow{7}{*}{10}
  & \RomanNum{1} & 2.78 & 2.78 & 0.95 & 0.94 & 0.77 & 0.76 & 0.97 & 0.96 \\ 
  & \RomanNum{2} & 2.70 & 2.73 & 0.93 & 0.95 & 0.71 & 0.72 & 0.89 & 0.97 \\ 
  & \RomanNum{3} & 2.59 & 2.61 & 0.95 & 0.95 & 0.73 & 0.74 & 0.88 & 0.96 \\ 
  & \RomanNum{4} & 2.89 & 2.98 & 0.88 & 0.94 & 0.71 & 0.72 & 0.63 & 0.98 \\ 
  & \RomanNum{5} & 2.90 & 2.90 & 0.95 & 0.95 & 0.75 & 0.75 & 0.76 & 0.96 \\ 
  & \RomanNum{6} & 3.98 & 4.01 & 0.93 & 0.95 & 0.79 & 0.79 & 0.97 & 0.99 \\ 
  & \RomanNum{7} & 3.22 & 3.27 & 0.86 & 0.94 & 0.77 & 0.77 & 0.77 & 1.00 \\ 
  \midrule
  \multirow{7}{*}{50}
  & \RomanNum{1} & 2.76 & 2.75 & 0.94 & 0.94 & 0.76 & 0.76 & 0.82 & 0.98 \\ 
  & \RomanNum{2} & 2.70 & 2.72 & 0.93 & 0.94 & 0.71 & 0.71 & 0.80 & 0.96 \\ 
  & \RomanNum{3} & 2.59 & 2.59 & 0.94 & 0.95 & 0.73 & 0.73 & 0.63 & 0.98 \\ 
  & \RomanNum{4} & 2.89 & 2.96 & 0.88 & 0.94 & 0.71 & 0.72 & 0.48 & 0.98 \\ 
  & \RomanNum{5} & 2.87 & 2.87 & 0.93 & 0.94 & 0.74 & 0.74 & 0.33 & 0.96 \\ 
  & \RomanNum{6} & 3.95 & 3.99 & 0.91 & 0.94 & 0.78 & 0.79 & 0.89 & 0.99 \\ 
  & \RomanNum{7} & 3.21 & 3.27 & 0.88 & 0.94 & 0.76 & 0.77 & 0.63 & 0.99 \\ 
   \bottomrule
\end{tabular*}
\end{table}

\section{Accuracy of Variable Selection}

Consider the simulated experiments in the main paper.
To quantify variable selection accuracy, we compute the true
positive rate, the number of signal variables included in the
decision list divided by the number of signal variables, and the false
positive rate, the number of noise variables included in the
decision list divided by the number of noise variables.
A variable is called a signal variable if it appears in
$\phi(x,a)$ and is a noise variable otherwise, irrespective of the actual functional form.

Table~\ref{tbl:varsel} presents the true positive rates and the false positive rates under different settings.
The proposed method consistently identifies signal variables and screens out noise variables in most settings.
The only exception is setting~\RomanNum{4}, where the optimal regime is far away from being well approximated
by decision lists. Thus the proposed approach loses power in detecting useful covariates due to
misspecifying the form of the regime.

\begin{table}
\renewcommand{\baselinestretch}{1.1}\fontsize{11}{12}\selectfont
\centering
\caption{Accuracy of variable selection using decision list.
TPR is the true positive rate and FPR is the false positive rate.}
\label{tbl:varsel}
\begin{tabular*}{\columnwidth}{c *{5}{@{\extracolsep{\fill}} c}}
  \toprule
 \multirow{2}{*}{\;$p$\;} & \multirow{2}{*}{Setting} & 
  \multicolumn{2}{c}{Continuous response} & \multicolumn{2}{c}{Binary response} \\
  \cmidrule{3-4} \cmidrule{5-6}
 & & TPR & FPR & TPR & FPR \\ 
  \midrule
  \multirow{7}{*}{10}
  & \RomanNum{1} & 1.00 & 0.00 & 1.00 & 0.07 \\ 
  & \RomanNum{2} & 1.00 & 0.00 & 1.00 & 0.04 \\ 
  & \RomanNum{3} & 1.00 & 0.00 & 1.00 & 0.11 \\  
  & \RomanNum{4} & 0.93 & 0.00 & 0.79 & 0.07 \\ 
  & \RomanNum{5} & 1.00 & 0.07 & 1.00 & 0.20 \\ 
  & \RomanNum{6} & 1.00 & 0.05 & 1.00 & 0.10 \\ 
  & \RomanNum{7} & 0.94 & 0.00 & 0.98 & 0.04 \\ 
  \midrule
  \multirow{7}{*}{50}
  & \RomanNum{1} & 1.00 & 0.01 & 1.00 & 0.04 \\ 
  & \RomanNum{2} & 1.00 & 0.00 & 1.00 & 0.02 \\ 
  & \RomanNum{3} & 1.00 & 0.00 & 0.99 & 0.04 \\ 
  & \RomanNum{4} & 0.93 & 0.00 & 0.73 & 0.02 \\ 
  & \RomanNum{5} & 1.00 & 0.02 & 0.99 & 0.06 \\ 
  & \RomanNum{6} & 1.00 & 0.02 & 1.00 & 0.03 \\ 
  & \RomanNum{7} & 0.94 & 0.00 & 0.97 & 0.02 \\ 
   \bottomrule
\end{tabular*}
\end{table}

\section{Impact of the Tuning Parameter in the Stopping Criterion}

In the algorithm discussed in Section~2.4.1 in the main paper, we use a tuning parameter $\alpha$
to control the building process of the decision list and we suggest to fix $\alpha$ at $0.95$. 
In the following we show that the final decision list is insensitive to the choice of $\alpha$ via simulation study.
The setting used here is exactly the same as that in Section~3 in the main paper.
We varied $\alpha$ among $\{0.9, 0.95, 0.99\}$.

Table~\ref{tbl:alpha-value} shows the impact of $\alpha$ on the value and the cost of the estimated regime.
We can see that the value and the cost as well as the accuracy of variable selection, averaged over 1000 replications,
are very stable across different choices of $\alpha$.
Table~\ref{tbl:alpha-regime} shows the impact of $\alpha$ on the estimated regime.
It is clear that $\alpha$ has little impact on the treatment recommendation made by the estimated regime.

\begin{table}
\renewcommand{\baselinestretch}{1.1}\fontsize{11}{12}\selectfont
\centering
\caption{The impact of $\alpha$ on the value and the cost of the estimated regime.
In the header, $\alpha$ is the tuning parameter in the stopping criterion;
$R(\wh{\pi})$ is the mean outcome under the estimated regime $\wh{\pi}$,
computed on a test set of $10^6$ subjects;
$N(\wh{\pi})$ is the cost of implementing the estimated regime $\wh{\pi}$,
computed on the same test set;
TPR is the true positive rate, namely, the number of signal variables involved in $\wh{\pi}$
divided by the number of signal variables;
FPR is the false positive rate, namely, the number of noise variables involved in $\wh{\pi}$
divided by the number of noise variables.
Recall that $p$ is the dimension of patient covariates.}
\label{tbl:alpha-value}
\begin{tabular*}{\columnwidth}{c @{\extracolsep{\fill}} c *{12}{ @{\extracolsep{\fill}} c}}
  \toprule
\multirow{2}{*}{$p$} & \multirow{2}{*}{Setting}
  & \multicolumn{4}{c}{$\alpha=0.9$}
  & \multicolumn{4}{c}{$\alpha=0.95$} 
  & \multicolumn{4}{c}{$\alpha=0.99$} \\
  \cmidrule{3-6} \cmidrule{7-10} \cmidrule{11-14}
&
  & $R(\widehat{\pi})$ & $N(\widehat{\pi})$ & TPR & FPR 
  & $R(\widehat{\pi})$ & $N(\widehat{\pi})$ & TPR & FPR 
  & $R(\widehat{\pi})$ & $N(\widehat{\pi})$ & TPR & FPR \\ 
  \midrule
\multicolumn{10}{l}{\emph{Continuous response}}\\
  \midrule
  \multirow{7}{*}{$10$}
  & \RomanNum{1} & 2.78 & 1.65 & 1.00 & 0.01 & 2.78 & 1.65 & 1.00 & 0.00 & 2.78 & 1.65 & 1.00 & 0.00 \\ 
  & \RomanNum{2} & 2.71 & 1.66 & 1.00 & 0.00 & 2.70 & 1.66 & 1.00 & 0.00 & 2.69 & 1.66 & 1.00 & 0.00 \\ 
  & \RomanNum{3} & 2.59 & 1.69 & 1.00 & 0.00 & 2.59 & 1.69 & 1.00 & 0.00 & 2.59 & 1.69 & 1.00 & 0.00 \\ 
  & \RomanNum{4} & 2.89 & 2.51 & 0.93 & 0.00 & 2.89 & 2.51 & 0.93 & 0.00 & 2.89 & 2.51 & 0.92 & 0.00 \\ 
  & \RomanNum{5} & 2.90 & 1.91 & 1.00 & 0.07 & 2.90 & 1.91 & 1.00 & 0.07 & 2.90 & 1.91 & 1.00 & 0.07 \\ 
  & \RomanNum{6} & 3.98 & 1.61 & 1.00 & 0.06 & 3.98 & 1.61 & 1.00 & 0.05 & 3.98 & 1.61 & 1.00 & 0.05 \\ 
  & \RomanNum{7} & 3.22 & 2.56 & 0.94 & 0.00 & 3.22 & 2.56 & 0.94 & 0.00 & 3.21 & 2.56 & 0.94 & 0.00 \\ 
   \midrule
  \multirow{7}{*}{$50$}
  & \RomanNum{1} & 2.75 & 1.96 & 1.00 & 0.01 & 2.76 & 1.96 & 1.00 & 0.01 & 2.78 & 1.96 & 1.00 & 0.00 \\ 
  & \RomanNum{2} & 2.70 & 1.66 & 1.00 & 0.00 & 2.70 & 1.66 & 1.00 & 0.00 & 2.69 & 1.66 & 1.00 & 0.00 \\ 
  & \RomanNum{3} & 2.58 & 1.75 & 1.00 & 0.00 & 2.59 & 1.75 & 1.00 & 0.00 & 2.59 & 1.75 & 1.00 & 0.00 \\ 
  & \RomanNum{4} & 2.89 & 2.54 & 0.93 & 0.00 & 2.89 & 2.54 & 0.93 & 0.00 & 2.89 & 2.54 & 0.92 & 0.00 \\ 
  & \RomanNum{5} & 2.87 & 2.19 & 1.00 & 0.03 & 2.87 & 2.19 & 1.00 & 0.02 & 2.88 & 2.19 & 1.00 & 0.02 \\ 
  & \RomanNum{6} & 3.95 & 1.70 & 1.00 & 0.02 & 3.95 & 1.70 & 1.00 & 0.02 & 3.95 & 1.70 & 1.00 & 0.02 \\ 
  & \RomanNum{7} & 3.22 & 2.56 & 0.94 & 0.00 & 3.21 & 2.56 & 0.94 & 0.00 & 3.21 & 2.56 & 0.93 & 0.00 \\ 
  \midrule
\multicolumn{10}{l}{\emph{Binary response}}\\
  \midrule
  \multirow{7}{*}{$10$}
  & \RomanNum{1} & 0.76 & 2.16 & 1.00 & 0.12 & 0.77 & 2.16 & 1.00 & 0.07 & 0.77 & 2.16 & 1.00 & 0.02 \\ 
  & \RomanNum{2} & 0.71 & 1.75 & 1.00 & 0.06 & 0.71 & 1.75 & 1.00 & 0.04 & 0.71 & 1.75 & 1.00 & 0.02 \\ 
  & \RomanNum{3} & 0.73 & 2.24 & 1.00 & 0.15 & 0.73 & 2.24 & 1.00 & 0.11 & 0.74 & 2.24 & 0.99 & 0.04 \\ 
  & \RomanNum{4} & 0.71 & 2.48 & 0.81 & 0.08 & 0.71 & 2.48 & 0.79 & 0.07 & 0.71 & 2.48 & 0.70 & 0.06 \\ 
  & \RomanNum{5} & 0.75 & 2.64 & 1.00 & 0.24 & 0.75 & 2.64 & 1.00 & 0.20 & 0.75 & 2.64 & 1.00 & 0.16 \\ 
  & \RomanNum{6} & 0.79 & 2.11 & 1.00 & 0.11 & 0.79 & 2.11 & 1.00 & 0.10 & 0.79 & 2.11 & 1.00 & 0.10 \\ 
  & \RomanNum{7} & 0.77 & 2.87 & 0.98 & 0.06 & 0.77 & 2.87 & 0.98 & 0.04 & 0.76 & 2.87 & 0.97 & 0.02 \\ 
  \midrule
  \multirow{7}{*}{$50$}
  & \RomanNum{1} & 0.75 & 2.87 & 1.00 & 0.05 & 0.76 & 2.87 & 1.00 & 0.04 & 0.76 & 2.87 & 1.00 & 0.02 \\ 
  & \RomanNum{2} & 0.71 & 1.93 & 1.00 & 0.02 & 0.71 & 1.93 & 1.00 & 0.02 & 0.71 & 1.93 & 0.99 & 0.01 \\ 
  & \RomanNum{3} & 0.72 & 2.68 & 0.99 & 0.04 & 0.73 & 2.68 & 0.99 & 0.04 & 0.73 & 2.68 & 0.99 & 0.02 \\ 
  & \RomanNum{4} & 0.71 & 2.65 & 0.75 & 0.02 & 0.71 & 2.65 & 0.73 & 0.02 & 0.71 & 2.65 & 0.66 & 0.02 \\ 
  & \RomanNum{5} & 0.73 & 3.32 & 0.99 & 0.07 & 0.74 & 3.32 & 0.99 & 0.06 & 0.74 & 3.32 & 0.99 & 0.05 \\ 
  & \RomanNum{6} & 0.78 & 2.47 & 1.00 & 0.03 & 0.78 & 2.47 & 1.00 & 0.03 & 0.78 & 2.47 & 1.00 & 0.02 \\ 
  & \RomanNum{7} & 0.76 & 3.04 & 0.97 & 0.02 & 0.76 & 3.04 & 0.97 & 0.02 & 0.76 & 3.04 & 0.95 & 0.01 \\ 
  \bottomrule
\end{tabular*}
\end{table}

\begin{table}
\renewcommand{\baselinestretch}{1.1}\fontsize{11}{12}\selectfont
\centering
\caption{The impact of $\alpha$ on the estimated regime.
In the header, $\alpha$ is the tuning parameter in the stopping criterion
and $\wh{\pi}_\alpha$ is the regime such obtained.
For each pair of regimes $\wh{\pi}_{\alpha}$ and $\wh{\pi}_{\alpha'}$,
we report the probability that they recommend the same treatment for a randomly selected patient in the population.
Mathematically, this is to compute $\pr\{\wh{\pi}_{\alpha}(X) = \wh{\pi}_{\alpha'}(X) | \wh{\pi}_{\alpha}, \wh{\pi}_{\alpha'}\}$
and then average over 1000 replications, where $X$ is generated in the same way as in Section~3 in the main paper.}
\label{tbl:alpha-regime}
\begin{tabular*}{\columnwidth}{c @{\extracolsep{\fill}} c *{3}{ @{\extracolsep{\fill}} c}}
  \toprule
\;$p$\; & Setting & $\wh{\pi}_{0.9}$ vs. $\wh{\pi}_{0.95}$ & $\wh{\pi}_{0.95}$ vs. $\wh{\pi}_{0.99}$ & $\wh{\pi}_{0.9}$ vs. $\wh{\pi}_{0.99}$ \\
  \midrule
\multicolumn{5}{l}{\emph{Continuous response}}\\
  \midrule
  \multirow{7}{*}{$10$}
  & \RomanNum{1} & 0.998 & 0.998 & 0.996 \\ 
  & \RomanNum{2} & 0.986 & 0.975 & 0.961 \\ 
  & \RomanNum{3} & 0.993 & 0.992 & 0.986 \\ 
  & \RomanNum{4} & 0.997 & 0.997 & 0.993 \\ 
  & \RomanNum{5} & 0.999 & 1.000 & 0.998 \\ 
  & \RomanNum{6} & 0.998 & 0.997 & 0.995 \\ 
  & \RomanNum{7} & 0.991 & 0.988 & 0.979 \\ 
   \midrule
  \multirow{7}{*}{$50$}
  & \RomanNum{1} & 0.984 & 0.985 & 0.970 \\ 
  & \RomanNum{2} & 0.986 & 0.977 & 0.962 \\ 
  & \RomanNum{3} & 0.988 & 0.989 & 0.976 \\ 
  & \RomanNum{4} & 0.998 & 0.996 & 0.994 \\ 
  & \RomanNum{5} & 0.993 & 0.995 & 0.988 \\ 
  & \RomanNum{6} & 0.997 & 0.997 & 0.993 \\ 
  & \RomanNum{7} & 0.991 & 0.989 & 0.980 \\ 
   \midrule
\multicolumn{5}{l}{\emph{Binary response}}\\
  \midrule
  \multirow{7}{*}{$10$}
  & \RomanNum{1} & 0.971 & 0.969 & 0.941 \\ 
  & \RomanNum{2} & 0.978 & 0.964 & 0.944 \\ 
  & \RomanNum{3} & 0.971 & 0.952 & 0.926 \\ 
  & \RomanNum{4} & 0.983 & 0.955 & 0.941 \\ 
  & \RomanNum{5} & 0.973 & 0.969 & 0.944 \\ 
  & \RomanNum{6} & 0.992 & 0.993 & 0.985 \\ 
  & \RomanNum{7} & 0.976 & 0.968 & 0.944 \\ 
   \midrule
  \multirow{7}{*}{$50$}
  & \RomanNum{1} & 0.973 & 0.946 & 0.920 \\ 
  & \RomanNum{2} & 0.980 & 0.958 & 0.942 \\ 
  & \RomanNum{3} & 0.971 & 0.947 & 0.925 \\ 
  & \RomanNum{4} & 0.985 & 0.962 & 0.947 \\ 
  & \RomanNum{5} & 0.965 & 0.939 & 0.913 \\ 
  & \RomanNum{6} & 0.985 & 0.985 & 0.969 \\ 
  & \RomanNum{7} & 0.974 & 0.955 & 0.930 \\ 
  \bottomrule
\end{tabular*}
\end{table}

\section{Chronic Depression Data}

In the application considered in Section~4.2 in the main paper,
we applied the proposed method to construct an interpretable and parsimonious
treatment regime.
We follow \citet{gunter2011variable} and \citet{zhao2012estimating},
and use the following $50$ covariates:
\begin{enumerate}[leftmargin=1.5em]
%10
\item Gender: 1 if female, 0 if male;
\item Race: 1 if white, 0 otherwise;
\item Marital status \RomanNum{1}: 1 if single, 0 otherwise;
\item Marital status \RomanNum{2}: 1 if married or living with someone, 0 otherwise;
\item Body mass index: continuous;
\item Age at screening: continuous;
\item Having difficulty in planning family activity: 1 if strongly agree, 2 if agree, 3 if disagree, 4 if strongly disagree;
\item Supporting each other in the family: 1 if strongly agree, 2 if agree, 3 if disagree, 4 if strongly disagree;
\item Having problems with primary support group: 1 if yes, 0 if no;
\item Having problems related to the social environment: 1 if yes, 0 if no;
% 20
\item Having occupational problems: 1 if yes, 0 if no;
\item Having economic problems: 1 if yes, 0 if no;
\item Receiving psychotherapy for current depression: 1 if yes, 0 if no or don't know;
\item Receiving medication for current depression: 1 if yes, 0 if no or don't know;
\item Having received psychotherapy for past depressions: 1 if yes, 0 if no or don't know;
\item Having received medication for past depressions: 1 if yes, 0 if no or don't know;
\item Number of major depressive disorder (MDD) episodes: 1 if one, 2 if two, 3 if at least three;
\item Length of current MDD episode (in years): continuous;
\item Age at MDD onset: continuous;
\item MDD severity: 1 if mild, 2 if moderate, 3 if severe;
% 30
\item MDD type \RomanNum{1}: 1 if melancholic, 0 otherwise;
\item MDD type \RomanNum{2}: 1 if atypical, 0 otherwise;
\item Number of dysthymia episodes: 0 if zero, 1 if one, 2 if at least two;
\item Generalized anxiety: 0 if absent or inadequate information, 1 if subthreshold, 2 if threshold;
\item Anxiety disorder (not otherwise specified): 0 if absent or inadequate information, 1 if subthreshold, 2 if threshold;
\item Panic disorder: 0 if absent or inadequate information, 1 if subthreshold, 2 if threshold;
\item Social phobia: 0 if absent or inadequate information, 1 if subthreshold, 2 if threshold;
\item Specific phobia: 0 if absent or inadequate information, 1 if subthreshold, 2 if threshold;
\item Obsessive compulsive: 0 if absent or inadequate information, 1 if subthreshold, 2 if threshold;
\item Body dysmorphic disorder: 0 if absent or inadequate information, 1 if subthreshold, 2 if threshold;
% 40
\item Post-traumatic stress disorder: 0 if absent or inadequate information, 1 if subthreshold, 2 if threshold;
\item Anorexia nervosa: 0 if absent or inadequate information, 1 if subthreshold, 2 if threshold;
\item Alcohol abuse: 0 if absent, 1 if abuse, 2 if dependent;
\item Drug abuse (including cannabis, stimulant, opioid, cocaine, hallucinogen): 0 if absent, 1 if abuse, 2 if dependent;
\item Global assessment of functioning: continuous;
\item Chronic depression diagnosis \RomanNum{1}: 1 if no antecedent dysthymia and continuous full-syndrome type;
\item Chronic depression diagnosis \RomanNum{2}: 1 if no antecedent dysthymia and incomplete recovery type;
\item Chronic depression diagnosis \RomanNum{3}: 1 if superimposed on antecedent dysthymia;
\item Chronic depression severity: integer between 1 (normal) and 7 (extremely ill);
\item Hamilton anxiety rating scale (HAM-A) total score: continuous;
%50
\item HAM-A psychic anxiety score: continuous;
\item HAM-A somatic anxiety score: continuous;
\item Hamilton depression rating scale (HAM-D) total score: continuous;
\item HAM-D anxiety/somatic score: continuous;
\item HAM-D cognitive disturbance score: continuous;
\item HAM-D retardation score: continuous;
\item HAM-D sleep disturbance: continuous;
\item Inventory of Depressive Symptoms - Self Report (IDS-SR) anxiety/arousal score: continuous;
\item IDS-SR general/mood cognition score: continuous;
\item IDS-SR sleep score: continuous.
\end{enumerate}

\section{Consistency of the decision list}

Since the consistency of the decision list is difficult to analyze theoretically,
we present some empirical evidence that the decision list tends to be consistent.
We follow the simulated experiments considered in Section~4 in the main paper but increase the sample size.
We consider settings \RomanNum{1} and \RomanNum{5} only as the optimal regime in other settings
cannot be representable as a decision list.

The sample sizes considered and the associated results are presented in Table~\ref{tbl:consistency}.
For continuous response, the proposed method correctly identifies the form and the important covariates for treatment decision.
As $n$ increases, the loss in value decreases and the probability of recommending the best treatment increases.
Also, the mean squared error of estimating the cutoff values decreases at the rate of $n^{-1}$.
Results for binary response is qualitatively similar. Nevertheless, 
we may need a even larger sample size for the asymptotics to work.
Therefore, the simulation results provides evidence that the proposed method is consistent.

\begin{table}
\renewcommand{\baselinestretch}{1.1}\fontsize{11}{12}\selectfont
\caption{Consistency of the decision list.
In the header, $n$ is the sample size; $p$ is the number of predictors.
Loss is $R(\pi^{\mathrm{opt}}) - R(\wh{\pi})$, namely, the difference between the the value under the estimated regime and the value under the optimal regime.
Pr(best) is $\pr\{\wh{\pi}(X) = \pi^{\mathrm{opt}}(X) | \wh{\pi} \}$, namely, the probability that the treatment recommended by the estimate regime coincides with the treatment recommended by the optimal regime.
Loss and Pr(best) are averaged over 1000 replications.
Correct is the proportion of $\wh{\pi}$ having the same form and covariates as $\pi^{\mathrm{opt}}$ among $1000$ replications;
MSE${}_1$ is the mean squared error of the estimated cutoff for $X_1$;
MSE${}_2$ is the mean squared error of the estimated cutoff for $X_2$.}
\label{tbl:consistency}
\begin{tabular*}{\columnwidth}{c @{\extracolsep{\fill}} c *{7}{ @{\extracolsep{\fill}} c}}
  \toprule
 Setting & $n$ & $p$ & Loss & Pr(best) & Correct & MSE${}_1 (\times n)$ & MSE${}_2 (\times n)$ \\ 
  \midrule
\multicolumn{8}{l}{\emph{Continuous response}}\\
  \midrule
  I & $10^4$ & $10$ & 0.0023 & 0.9982 & 1.00 & 4.24 & 6.87 \\ 
  I & $10^5$ & $10$ & 0.0006 & 0.9995 & 1.00 & 4.30 & 6.50 \\ 
  I & $10^6$ & $10$ & 0.0002 & 0.9998 & 1.00 & 4.06 & 6.32 \\ 
  I & $10^4$ & $50$ & 0.0022 & 0.9982 & 1.00 & 4.60 & 6.91 \\ 
  I & $10^5$ & $50$ & 0.0006 & 0.9995 & 1.00 & 4.24 & 6.49 \\ 
  I & $10^6$ & $50$ & 0.0002 & 0.9998 & 1.00 & 4.22 & 6.48 \\ 
  \midrule
  V & $10^4$ & $10$ & 0.0039 & 0.9975 & 1.00 & 6.08 & 5.27 \\ 
  V & $10^5$ & $10$ & 0.0010 & 0.9994 & 1.00 & 5.86 & 4.70 \\ 
  V & $10^6$ & $10$ & 0.0003 & 0.9998 & 1.00 & 5.46 & 4.54 \\ 
  V & $10^4$ & $50$ & 0.0036 & 0.9977 & 1.00 & 6.10 & 5.33 \\ 
  V & $10^5$ & $50$ & 0.0010 & 0.9994 & 1.00 & 5.96 & 4.54 \\ 
  V & $10^6$ & $50$ & 0.0003 & 0.9998 & 1.00 & 5.69 & 4.51 \\ 
  \midrule
\multicolumn{8}{l}{\emph{Binary response}}\\
  \midrule
  I & $10^4$ & $10$ & 0.0007 & 0.9966 & 1.00 & 8.13 & 10.93 \\ 
  I & $10^5$ & $10$ & 0.0001 & 0.9994 & 1.00 & 5.71 & 6.05 \\ 
  I & $10^6$ & $10$ & 0.0000 & 0.9999 & 1.00 & 5.27 & 5.61 \\ 
  I & $10^4$ & $50$ & 0.0007 & 0.9965 & 1.00 & 9.72 & 11.15 \\ 
  I & $10^5$ & $50$ & 0.0001 & 0.9994 & 1.00 & 5.71 & 6.29 \\ 
  I & $10^6$ & $50$ & 0.0000 & 0.9998 & 1.00 & 5.41 & 5.78 \\ 
  \midrule
  V & $10^4$ & $10$ & 0.0094 & 0.9447 & 0.79 & 7.81 & 22.66 \\ 
  V & $10^5$ & $10$ & 0.0081 & 0.9547 & 0.96 & 4.14 & 6.33 \\ 
  V & $10^6$ & $10$ & 0.0079 & 0.9563 & 0.97 & 3.89 & 5.03 \\ 
  V & $10^4$ & $50$ & 0.0099 & 0.9418 & 0.70 & 6.63 & 14.36 \\ 
  V & $10^5$ & $50$ & 0.0078 & 0.9567 & 0.96 & 3.97 & 6.14 \\ 
  V & $10^6$ & $50$ & 0.0081 & 0.9550 & 0.97 & 3.96 & 5.10 \\ 
  \bottomrule
\end{tabular*}
\end{table}

\end{document}